\documentclass[useAMS,usenatbib,usegraphicx]{mn2e}
\usepackage{ gensymb }

\usepackage[fleqn]{amsmath}
\usepackage{amssymb}

\newcommand{\dd}{{\rm d}}
\newcommand{\Pm}{\mathcal{P}}

\newcommand\rmd{\mathrm{d}}

\newcommand\rms{\mathrm{s}}

\newcommand\f{\frac}
\newcommand\p{\partial}

\usepackage{epsfig}
\usepackage{amsmath}

\title[Magnetic flux transport in accretion discs]
{Transport of magnetic flux and the vertical structure of accretion discs: II. Vertical profile of the diffusion coefficients}

\author[J\'er\^ome Guilet and Gordon I. Ogilvie]
{J\'er\^ome Guilet and Gordon I. Ogilvie\\
Department of Applied Mathematics and Theoretical Physics,
University of Cambridge, Centre for Mathematical Sciences,\\
Wilberforce Road, Cambridge CB3 0WA
}

\begin{document}

\maketitle

\label{firstpage}

\begin{abstract}
We investigate the radial transport of magnetic flux in a thin accretion disc, the turbulence being modelled by effective diffusion coefficients (viscosity and resistivity). Both turbulent diffusion and advection by the accretion flow contribute to flux transport, and they are likely to act in opposition. We study the consequences of the vertical variation of the diffusion coefficients, due to a varying strength of the turbulence. For this purpose, we consider three different vertical profiles of these coefficients. The first one is aimed at mimicking the turbulent stress profile observed in numerical simulations of MHD turbulence in stratified discs. This enables us to confirm the robustness of the main result of Paper~I obtained for uniform diffusion coefficients that, for weak magnetic fields, the contribution of the accretion flow to the transport velocity of magnetic flux is much larger than the transport velocity of mass. We then consider the presence of a dead zone around the equatorial plane, where the physical resistivity is high while the turbulent viscosity is low. We find that it amplifies the previous effect: weak magnetic fields can be advected orders of magnitude faster than mass, for dead zones with a large vertical extension. The ratio of advection to diffusion, determining the maximum inclination of the field at the surface of the disc, is however not much affected. Finally, we study the effect of a non-turbulent layer at the surface of the disc, which has been suggested as a way to reduce the diffusion of the magnetic flux. We find that the reduction of the diffusion requires the conducting layer to extend below the height at which the magnetic pressure equals the thermal pressure. As a consequence, if the absence of turbulence is caused by the large-scale magnetic field, the highly conducting layer is inefficient at reducing the diffusion.
\end{abstract}


\section{Introduction}
Jets or outflows are observed in a variety of objects where an accretion disc rotates and slowly accretes onto a central object. This central object varies considerably in mass from a supermassive black hole in AGNs, to a stellar-mass black hole or a neutron star in X-ray binaries, a white dwarf in cataclysmic variables or a pre-main sequence star in T~Tauri star systems. The presence of outflows in all these objects at very different scales suggests a deep connection between the accretion process through an accretion disc and the launching process of the outflow, which is further suggested by a correlation between the observational signatures of ejection and accretion. Theoretical models generally appeal to a large-scale magnetic field to accelerate and collimate the outflow into a jet. The magnetic field serves as an agent to extract the energy contained in the rotation of the accretion disc in the magnetocentrifugal mechanism \citep{blandford82} or of the central black hole \citep{blandford77}. It also has the advantage of providing a natural collimation of the outflow. But can an accretion disc possess a strong magnetic field as required by these models? And what is its origin?

One possibility is a dynamo arising from the magnetorotational instability (MRI, \citet{balbus91}). However, the dynamo process would need to be very efficient to produce the right strength of the magnetic field. It also remains unclear whether a {\it large-scale} magnetic field can be created in this way, because the correlation length of the turbulence is probably of the order of the vertical scale height of the disc, which is much smaller than its radius (see the discussion in
\citealt{spruit10}). 

In another scenario, the gas supplied to the disc contains an initially weak magnetic field which is transported inwards by the accretion flow. The magnetic field is strongly amplified as magnetic flux accumulates in the central part of the accretion disc, and could thus reach the required strength to launch an outflow. The efficiency of this scenario depends upon the relative efficiency of the inward advection of magnetic flux by the accretion flow and of the diffusion of the magnetic field by the action of the effective turbulent resistivity, which tends to counteract the advection process and transport magnetic flux outwards. Unfortunately, early theoretical studies suggested that in thin accretion discs the diffusion process is much more efficient than the advection process \citep{lubow94a,heyvaerts96}. Indeed the diffusion arises from the vertical diffusion of the radial component of the magnetic field across the accretion disc. In a thin disc, this takes place on a much smaller scale than the radius, which is the scale on which the effective viscosity transports angular momentum thus driving the advection. This scale difference makes the diffusion much more efficient than the diffusion. This raises doubts over the existence of a significant large-scale magnetic field in accretion discs and is therefore a major problem for models of jets. 

The aforementioned results were, however, obtained from a crude vertical averaging, and it was realised later that the vertical structure of the accretion disc might change the picture \citep{ogilvie01,bisnovatyi-kogan07,guilet12}. The purpose of this series of articles is to investigate the influence of the vertical structure of the disc on the transport of magnetic flux, the effect of turbulence being modelled by effective diffusion coefficients (viscosity and resistivity). \citet{guilet12} (hereafter called Paper~I) was restricted to the idealised case of uniform diffusion coefficients. It was found that even in that simple case the vertical profiles of the velocity and magnetic field have a significant impact on the transport of magnetic flux (particularly for magnetic fields that are weak in the sense that the magnetic pressure is small compared to the thermal pressure in the equatorial plane). This article is devoted to studying the impact of a vertical variation of the diffusion coefficients. It is indeed expected that the turbulence strength varies with the distance from the midplane, as shown by numerical simulations of stratified discs \citep[e.g.][]{miller00,fromang06b}. This should result in a vertical variation of the effective diffusion coefficients. Three different physically motivated profiles are considered in this paper:
\begin{itemize}
\item The first one, inspired by MHD simulations of fully turbulent discs, is mostly aimed at testing the robustness of the results obtained in Paper~I with a more realistic description of the vertical structure of the turbulence. 
\item Secondly, we consider the presence of a dead zone close to the equatorial plane, where the (physical) resistivity is large owing to low ionisation. As a consequence the MRI-induced turbulence is switched off, resulting in a small effective viscosity. Dead zones are expected to exist in some parts of protoplanetary discs \citep{gammie96,fromang02} and could therefore influence the transport of magnetic field in these objects. 
\item Finally, a third profile of the diffusion coefficients is considered in order to test the scenario proposed by \citet{bisnovatyi-kogan07} to reduce the diffusion of the magnetic flux. It relies on a non-turbulent layer at the surface of the disc, where the magnetic field is strong enough to stabilise the MRI. According to this scenario, the reduced turbulent diffusion in this layer would allow the magnetic field lines to bend without much diffusion. However, our self-consistent calculation of the vertical structure with a non-turbulent surface layer shows that this layer is inefficient at reducing the diffusion unless it extends to where the magnetic field is {\it not} expected to stabilise the MRI.
\end{itemize}

In Section~\ref{sec:formalism}, we summarise the formalism developed in Paper~I and used in the rest of the paper to compute the vertical structure of an accretion disc and the transport rate of magnetic flux. In Section~\ref{sec:cst_stress}, we consider a vertical profile of the turbulent viscosity that mimics the results of MHD simulations of fully turbulent discs. In Section~\ref{sec:dead_zone}, we model a non-turbulent dead zone around the equatorial plane. Finally in Section~\ref{sec:lovelace}, we study the effect of a non-turbulent surface layer on the diffusion of the magnetic flux. We discuss the results and conclude in Section~\ref{sec:conclusion}.

\section{Formalism}
	\label{sec:formalism}
\subsection{Assumptions and governing equations}

In Paper~I, we derived a formalism enabling us to compute the transport of magnetic flux while taking into account the vertical structure of an accretion disc. In this section we summarise the approach and give the governing equations; the reader is referred to Paper~I for more details on their derivation. The formalism is based on a systematic asymptotic expansion in the limit of a thin accretion disc. For consistency in the expansion, as well as to obtain a linear system of equations which can be more easily solved, we further assume that the magnetic field is dominated by its vertical component. This amounts to neglecting the compression of the disc due to magnetic pressure. While this is an important limitation for strong magnetic fields, we showed in Paper~I by comparing with direct numerical simulations that this approximation gives good results for the range of magnetic field strengths considered in this paper. We assume a point-mass potential and therefore circular Keplerian orbital motion at leading order. An isothermal equation of state is assumed for simplicity. We look for solutions that are stationary on a dynamical timescale but may evolve on the longer viscous or resistive timescales. 

In order to model the effects of turbulence, we use effective turbulent diffusion coefficients: a viscosity $\nu$ and a magnetic diffusivity (proportional to resistivity) $\eta$. The standard alpha prescription is used for the viscosity: $\nu = \alpha c_\mathrm{s} H$, where $c_\mathrm{s}$ is the sound speed and $H$ is the vertical scaleheight $H\equiv c_\mathrm{s}/\Omega_\mathrm{K}$ with $\Omega_\mathrm{K}$ the Keplerian angular velocity. We allow for a vertical (but not radial) dependence of the $\alpha$ parameter through $\alpha = \alpha_0g(\zeta)$. The resistivity is then related to the viscosity through the magnetic Prandtl number $\Pm \equiv \nu/\eta $, the vertical profile of which can also be freely chosen. The normalisation of the diffusion coefficients (i.e. the value of $\alpha_0$) is determined by requiring that the MRI is made marginally stable by the diffusion. This ensures that the stationary solutions found are not unstable, and prevents unphysical effects such as multiple bending of the magnetic field lines that arise when unstable MRI modes exist (see Paper~I for a discussion of this assumption). As already mentioned, the vertical profile of the diffusion coefficients can be freely chosen, and three different profiles are considered in Sections~\ref{sec:cst_stress}, \ref{sec:dead_zone} and \ref{sec:lovelace}. 

We use dimensionless variables defined as
\begin{eqnarray}
\tilde{\rho} &\equiv & \frac{\rho H}{\Sigma}, \\
u_r &\equiv & \frac{r}{H}\frac{v_{r}}{c_\rms}, \\
u_\phi &\equiv &  \frac{r}{H}\frac{v_{\phi}-v_\mathrm{K}}{c_\rms}, \\
b_r &\equiv &  \frac{r}{H}\frac{B_{r}}{B_z}, \\
b_\phi &\equiv &  \frac{r}{H}\frac{B_{\phi}}{B_z}, 
\end{eqnarray}
where $\rho$ is the gas density, $\Sigma$ the surface density, $v_\mathrm{K}=r\Omega_\mathrm{K}$ the Keplerian azimuthal velocity, $v_r$ and $v_\phi$ the radial and azimuthal components of the velocity, and $B_r$ and $B_\phi$ the radial and azimuthal components of the magnetic field. The vertical spatial coordinate $z$ is normalised by the disc scale height, defining
\begin{equation}
  \zeta \equiv \frac{z}{H}.
\end{equation}The vertical component of the magnetic field $B_z$ does not depend on $z$ in the limit of a thin disc and is therefore a free parameter of the problem. In the absence of an outflow due to the low inclination of the magnetic field lines, the vertical component of the velocity is negligibly small.

With the isothermal equation of state, the density profile is a Gaussian function:
\begin{equation}
\tilde{\rho} = \frac{1}{\sqrt{2\pi}}\,\mathrm{e}^{-\zeta^2/2}.
\end{equation}

Under these assumptions, the vertical structure of $u_r$, $u_\phi$, $b_r$ and $b_\phi$ is determined by a quasi-local effectively stationary problem, where global gradients appear as source terms. The governing differential system is 
\begin{eqnarray}
-\frac{1}{\tilde\rho}\lefteqn{\p_\zeta\left(\tilde{\rho}\alpha\p_\zeta u_r \right) - 2u_\phi - \frac{1}{\beta_0\tilde\rho} \p_\zeta b_r = \frac{3}{2} + D_H - D_{\nu\Sigma}}&\nonumber\\
&& + \left(\frac{3}{2}-D_H\right)\zeta^2 - \frac{D_B}{\beta_0\tilde\rho}, \label{eq:motion_r}
\end{eqnarray}
\begin{eqnarray}
-\frac{1}{\tilde\rho}\lefteqn{\p_\zeta\left(\tilde{\rho}\alpha\p_\zeta u_\phi \right) + \frac{1}{2} u_r - \frac{1}{\beta_0\tilde\rho}\p_\zeta b_\phi = \frac{3}{2}\alpha \bigg\lbrack - \frac{1}{2} }&\nonumber\\
&& + D_H\left(1 + \frac{\dd\ln\alpha}{\dd\ln\zeta} \right)- D_{\nu\Sigma} - D_H\zeta^2 \bigg\rbrack,  \label{eq:motion_phi}
\end{eqnarray}
\begin{equation}
-\p_\zeta\left( \frac{\alpha}{\Pm}\p_\zeta b_r \right) - \p_\zeta
u_r = -D_B \p_\zeta\left(\frac{\alpha}{\Pm}\right), \label{eq:induction_r}
\end{equation}
\begin{equation}
 -\p_\zeta\left(\frac{\alpha}{\Pm}\p_\zeta b_\phi\right) - \p_\zeta u_\phi + \frac{3}{2}b_r = 0,\label{eq:induction_phi}
\end{equation}
where we have defined the following dimensionless parameters:
\begin{eqnarray}
\beta_0 &\equiv& \frac{\mu_0}{B_z^2}\frac{\Sigma c_\mathrm{s}^2}{H}, \\
D_H &\equiv&\frac{\p\ln H}{\p\ln r}, \\
D_{\nu\Sigma} &\equiv& 2D_H - \frac{3}{2} + \frac{\p\ln\Sigma}{\p\ln r}, \\
D_B &\equiv & \frac{\p\ln B_z}{\p\ln r}.
\end{eqnarray}
Here $\beta_0$ corresponds roughly to the midplane value of the plasma $\beta$ parameter (the ratio of the thermal pressure to the magnetic pressure); more precisely, the two are related by $\beta(\zeta=0) = \sqrt{2/\pi}\,\beta_0 $. The parameter $D_{\nu\Sigma}$ equals $\p\ln(\nu\Sigma)/\p\ln r$, given that $\nu=\alpha H^2\Omega$ and $\alpha$ does not depend on $r$.

The boundary conditions at $\zeta\rightarrow\pm\infty$ are motivated by the absence of outflow, owing to the low inclination of the field lines. As a consequence the magnetic field at infinity is force-free, giving the following conditions at $\zeta \rightarrow \pm \infty$:
\begin{equation}
\rho u_r \to0, \label{eq:boundary_ur} \\
\end{equation}
\begin{equation}
\rho u_\phi \to0, \label{eq:boundary_uphi}
\end{equation}
\begin{equation}
b_r-(D_B\zeta\pm b_{r\rms})\to0, \label{eq:boundary_br}
\end{equation}
\begin{equation}
b_\phi - (\pm b_{\phi\rms}) \to 0. \label{eq:boundary_bphi} \\
\end{equation}
 The inclusion of a non-vanishing azimuthal component of the magnetic field as $\zeta \rightarrow \pm \infty$ allows for the possibility of a magnetic torque between the disc and the external medium.  If the disc is isolated and has no outflow, then $b_{\phi\rms}$ should be zero; however, the parameter $b_{\phi\rms}$ allows us, if we wish, to mimic in a simple way the effect of angular momentum removal by an outflow or by the connection to an external medium. These boundary conditions are homogeneous, except for the linear source terms proportional to $D_B$, $b_{r\rms}$, and $b_{\phi\rms}$ in equations~(\ref{eq:boundary_br}) and (\ref{eq:boundary_bphi}).

\subsection{Form of the solution}
 The system of equations given above is a linear system of ordinary differential equations for $u_r(\zeta)$, $u_\phi(\zeta)$, $b_r(\zeta)$ and $b_\phi(\zeta)$; radial derivatives appear only as parameters of this quasi-local problem. As a consequence of the linearity of the equations the solution depends linearly on the source terms, appearing either on the right-hand side of the system of equations or as a non-vanishing boundary condition at infinity ($b_{r\rms}$ and  $b_{\phi\rms}$). Thus, the general solution $\bmath{X}=\{u_r,u_\phi,b_r,b_\phi\}$ is a linear combination of the solution vectors corresponding to each source term:
\begin{eqnarray}
\lefteqn{\bmath{X} = \bmath{X}_\mathrm{K} + \bmath{X}_{DH}D_H + \bmath{X}_{D\nu\Sigma}D_{\nu\Sigma} + \bmath{X}_{DB}D_B}&\nonumber\\
&& + \bmath{X}_{br\rms}b_{r\rms} + \bmath{X}_{b\phi\rms}b_{\phi\rms},
\end{eqnarray}
where $\bmath{X}_{DH}$ is the solution vector corresponding to the source term proportional to $D_H$ and so on. $\bmath{X}_\mathrm{K}$ is the solution vector corresponding to the source terms $\bmath{F}$ that are independent of
$D_H$, $D_{\nu\Sigma}$, $D_B$, $b_{r\rms}$, and $b_{\phi\rms}$; these terms contain
the radial derivative of the leading-order angular velocity which is
assumed to be Keplerian (hence the notation), as well as a term describing the vertical dependence of the gravitational potential and other geometrical terms coming from differentiating factors depending on $r$ in the viscous term and the radial pressure gradient. Furthermore, we define
\begin{equation}
\bmath{X}_\mathrm{hyd} = \bmath{X}_\mathrm{K} + \bmath{X}_{DH},
\end{equation}
being the solution corresponding to the hydrodynamic source terms with
the standard parameters $D_H=1$ (relevant to a disc with a constant
aspect ratio $H/r$) and $D_{\nu\Sigma}=0$ (relevant to a steady
accretion disc far from the inner boundary). For these parameters, we
thus have
\begin{equation}
\bmath{X} = \bmath{X}_\mathrm{hyd} +  \bmath{X}_{DB}D_B + \bmath{X}_{br\rms}b_{r\rms} +  \bmath{X}_{b\phi\rms}b_{\phi\rms}.
\end{equation}

The linearity of the problem makes the exploration of the parameter
space and the physical understanding of the solutions much easier.
Indeed, given the marginal stability hypothesis, the solution depends
in a nonlinear way only on the two parameters $\beta_0$ and $\Pm$ (which is assumed to be unity in this article, except inside the dead zone as described in Section~\ref{sec:dead_zone}). For each pair of values of these two parameters, one needs to compute the
six solution vectors $\bmath{X}_\mathrm{K}$, $\bmath{X}_{DH}$, $
\bmath{X}_{D\nu\Sigma}$, $\bmath{X}_{DB}$,  $\bmath{X}_{br\rms}$ and $\bmath{X}_{b\phi\rms}$,
each of which can be represented by plotting the profiles of $u_r$,
$u_\phi$, $b_r$ and $b_\phi$.  The general solution is then just a
linear combination of these solution vectors with the appropriate
coefficients. To solve the problem numerically, we use a spectral method based on a decomposition on a basis of Whittaker functions as described in Paper~I.

\subsection{Transport velocities from vertical averaging}
	\label{sec:induction_average}

Once the vertical profiles of the velocity and magnetic field have been computed, the transport rates of mass and magnetic flux can be obtained and expressed in terms of a transport velocity. These transport rates determine the evolution of the disc on a viscous or resistive timescale. The dimensionless mass transport velocity is, from equation~(\ref{eq:motion_phi}),
\begin{equation}
  u_m=\int_{-\infty}^\infty\tilde\rho u_r\,\rmd\zeta =-\f{3}{2}\bar\alpha(1+2D_{\nu\Sigma}) + \frac{4}{\beta_0}b_{\phi\rms},
\label{um}
\end{equation}
where $\bar\alpha$ is the density-weighted vertically averaged dimensionless viscosity:
\begin{equation}
 \bar\alpha =  \int_{-\infty}^\infty \tilde{\rho}\alpha(\zeta)\,\rmd\zeta   \label{eq:density_av_alpha}.
\end{equation}
For a vanishing $b_{\phi\rms}$ this is consistent with the familiar expression from the standard theory of a Keplerian accretion disc,
\begin{equation}
  \bar v_r=-\f{3}{r^{1/2}\Sigma}\p_r(r^{1/2}\bar\nu\Sigma),
\end{equation}
where the overbar refers to a density-weighted vertical average.

The integrated version of the radial component~(\ref{eq:induction_r}) of the induction equation is
\begin{equation}
u_\psi = u_r + \frac{\alpha}{\Pm}(\p_\zeta b_r-D_B) = \mathrm{const},
\label{upsi}
\end{equation}
where
\begin{equation}
  u_\psi\equiv\f{r}{H}\f{v_\psi}{c_\rms}
\end{equation}
is the dimensionless magnetic flux transport velocity. It can be evaluated using equation~(\ref{upsi}) at any convenient height. In particular, the force-free conditions at $\zeta\rightarrow\pm\infty$ ensure that the radial velocity at this location equals $u_\psi$. This is useful to remember for the visual inspection of the vertical profiles displayed in Figures~\ref{fig:expalpha_brs0}, \ref{fig:expalpha_hydro0}, \ref{fig:dead_zone_brs}, \ref{fig:dead_zone_hydro} and \ref{fig:cutalpha_brs0}. A useful way of vertically averaging the induction equation to obtain the magnetic flux transport velocity is the following. Dividing equation~(\ref{upsi}) by the dimensionless resistivity $\alpha/\Pm$,
and integrating between $\zeta=0$ and
$\zeta$, we obtain:
\begin{equation}
 u_{\psi} = \bar{\eta}\left(\frac{b_r}{\zeta} - D_B \right) + \bar{u}_r,
 	\label{eq:upsi_average}
\end{equation}
where $\bar{\eta}$ is an average dimensionless resistivity defined in the following way (it is
actually the inverse of the height-averaged conductivity):
\begin{equation}
\bar{\eta} = \frac{\zeta}{\int_0^\zeta \frac{\Pm}{\alpha}\, \dd
  \zeta^{\prime}},   \label{eq:eta_average}
\end{equation}
and $\bar{u}_r$ is the average of the radial velocity weighted by the conductivity ($1/\eta$):
\begin{equation}
\bar{u}_r = \frac{1}{\int_0^\zeta \frac{\Pm}{\alpha}\, \dd \zeta^{\prime}}
\int_0^\zeta \frac{\Pm}{\alpha} u_r\, \dd \zeta^{\prime}.   \label{eq:vadv_average}
\end{equation}

This averaging procedure shows that the contribution of advection to the velocity at which
the magnetic flux is transported is the vertically averaged radial
velocity weighted by the electrical conductivity.  This average may be
very different from the density-weighted average $u_m$ that describes mass transport \citep{ogilvie01}.

Note that equation~(\ref{eq:upsi_average}) is valid for any height $\zeta$ up to which the averaging procedure is performed, and which can be chosen in the most convenient way. Integrating up to an infinite height may not be very useful as the force-free regime would dominate the average and one would simply recover that the radial velocity at infinity equals $u_\psi$. We showed in Paper~I that it is more meaningful to perform the average up to the height $\zeta_B$, defined as the height at which the magnetic and thermal pressures are equal:
\begin{equation}
\zeta_B = \sqrt{\ln\left(\frac{2}{\pi}\beta_0^2\right)}. 
	\label{eq:zetab}
\end{equation}

Indeed, we showed in Paper~I that for large enough values of $\beta_0$ ($\gtrsim 10^3$) the vertical profiles can be described by a two-zone model, where the magnetic field is approximately passive for $\zeta<\zeta_B$ and force-free for $\zeta>\zeta_B$. The velocity at $\zeta<\zeta_B$ is not influenced by the magnetic field, implying that $\bar{u}_r$ can be obtained from the purely hydrodynamical problem. The value of $b_r$ at $\zeta=\zeta_B$ is a bit less straightforward to obtain because $b_r$ can have a jump at the transition with the force-free region, where the boundary conditions at infinity can be applied. In Paper~I, we found that for the source terms governing the diffusion of the magnetic field ($b_{r\rms}$ and $D_B$) the value of $b_r$ at $\zeta=\zeta_B^-$ was a factor $3/2$ larger than that at $\zeta_B^+$ dictated by the boundary conditions: $b_r(\zeta_B^-) = (3/2)(b_{r\rms}+D_B\zeta_B)$. We find in the following sections that this numerical factor can depend on the vertical profile of the diffusion coefficients but remains of order unity.

\section{Model of a fully turbulent disc}
	\label{sec:cst_stress}

\begin{figure*}
   \centering
   \includegraphics[width=\columnwidth]{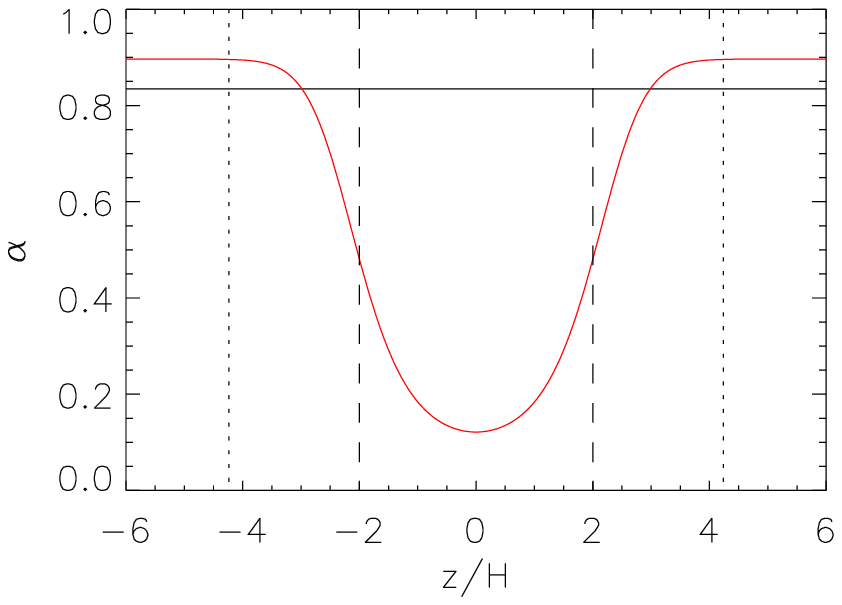}
    \includegraphics[width=\columnwidth]{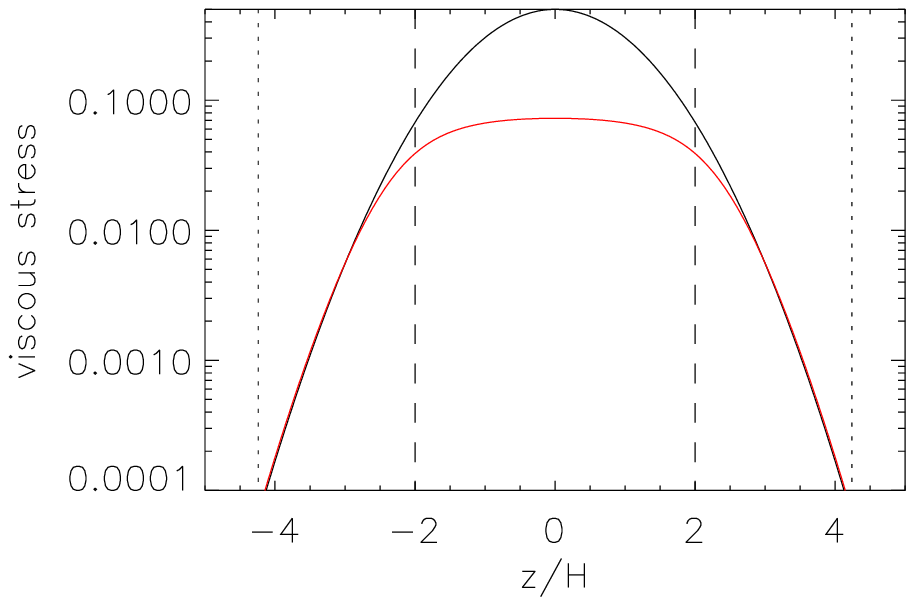}
   \caption{Model of a fully turbulent disc: vertical profiles of the dimensionless viscosity parameter $\alpha$ (left panel), and of the viscous stress ($r\phi$ component of the viscous stress tensor: $(3/2)\tilde{\rho}\alpha$, right panel) for a magnetic field strength $\beta_0=10^4$. The red line shows the profile used in Section~\ref{sec:cst_stress}  (equation~\ref{eq:exp_alpha}), while the black line shows the case of uniform diffusion coefficients studied in Paper~I. The vertical dashed lines show the height $\zeta_{c}$ below which the viscous stress is approximately uniform, and the dotted lines the height $\zeta_B$ above which the magnetic field is dominant.}
   \label{fig:shape_alpha_exp}
\end{figure*}

\begin{figure}
   \centering
   \includegraphics[width=\columnwidth]{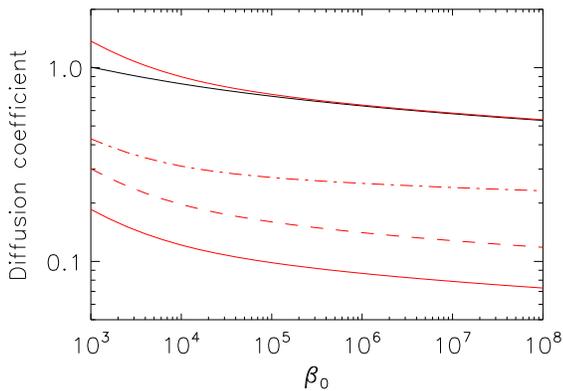}
   \caption{Diffusion coefficients determined by the marginal stability
     hypothesis as a function of the magnetic field strength. The black
     line correspond to a uniform $\alpha$ as studied in Paper~I. Red lines correspond to the
     profile described by equation~(\ref{eq:exp_alpha}): the full lines show the minimum and
     maximum values of $\alpha$,  the dashed line shows the density-weighted average defined by equation~(\ref{eq:density_av_alpha}) (relevant to the advection of mass), and the dash-dotted line shows the average resistivity defined by equation~(\ref{eq:eta_average}) with the average being performed up to $\zeta=\zeta_B$ (relevant to the diffusion of the magnetic field).}
   \label{fig:alpha_beta_exp}
\end{figure}

As a consequence of the vertical structure of the density, the properties of the turbulence are not expected to be uniformly distributed in the vertical direction inside an accretion disc. A large number of numerical simulations of 3D MHD turbulent discs with vertical stratification have been performed with both local and global models and can serve as a guide to build a more realistic model than the uniform diffusion coefficients model of Paper~I. Most of these simulations use a magnetic field configuration with zero net vertical flux because a significant net flux tends to cause numerical difficulties with the vertical boundaries in the presence of vertical stratification. One might, however, expect the stress profile in the case of a weak vertical field to be similar to that of these zero-net-flux calculations (at least close to the midplane, as long as $\beta \gg 1$). Numerical simulations of stratified discs with zero net vertical magnetic flux show that the disc is divided into two regions: the weakly magnetised main body of the disc at $|z| < 2-3H$, and a strongly magnetised corona at $|z| >2-3H$. In the main body of the disc the turbulent stress is roughly independent of the distance from the midplane \citep[e.g.][]{miller00,fromang06}. In the strongly magnetised corona the stress then decreases like the density \citep{miller00}. Interpreting this vertical profile of the stress as a vertical dependence of the $\alpha$ viscosity leads to a profile of $\alpha$ that is inversely proportional to the density for $|\zeta|<2-3$ and constant above $|\zeta|=2-3$. Such a vertical profile of $\alpha$ is probably more realistic than the uniform one for $\beta_0 \gg 1$. We choose the following functional dependence of $\alpha$:
\begin{equation}
\alpha(\zeta) =
\alpha_0\frac{e^{\zeta_c^2/2}}{1+(e^{\zeta_c^2/2}-1)e^{-\zeta^2/2}},
\label{eq:exp_alpha}
\end{equation}
where $\alpha_0$ is the value of $\alpha$ at the midplane, and the parameter $\zeta_c$ determines the height of the magnetic corona where $\alpha$ transitions from a profile approximately proportional to $e^{\zeta^2/2}$ (for $\zeta<\zeta_c$) to a constant value ($\alpha \simeq \alpha_0e^{\zeta_c^2/2}$ for $\zeta > \zeta_c$). In the following we use the value $\zeta_c=2$, corresponding to a vertical variation of $\alpha$ by a factor of $\alpha_\mathrm{max}/\alpha_0 = e^2 \simeq 7.4$. 

Figure~\ref{fig:shape_alpha_exp} shows the vertical profile of the $\alpha$ parameter (left panel) as well the viscous stress (right panel) obtained for our fiducial value of the magnetic field strength $\beta_0=10^4$. The $\alpha$ parameter ranges from $0.12$ at the midplane to $0.9$ in the corona. The density-weighted average of $\alpha$ (relevant to advection of mass) is $\bar{\alpha}=0.2$, while the average resistivity as defined in equation~(\ref{eq:eta_average}) (relevant to the diffusion of the magnetic flux) is $\bar{\eta}= 0.31$. The dependence of these various measures of the diffusion with respect to the magnetic field strength is shown in Figure~\ref{fig:alpha_beta_exp}. Note that for $\beta_0>10^4-10^5$ the value of $\alpha$ in the corona is the same as that obtained in the case of uniform diffusion coefficients (and ranges from $0.5$ for weak fields to $1$). This is explained by the fact that in the marginal stability hypothesis the value of $\alpha$ is determined by the largest-scale MRI channel modes, localised around $\zeta=\zeta_B$, which is in the corona since $\zeta_B>\zeta_c$. As a result of the profile considered, the midplane value as well as the vertical averages of the diffusion coefficients are significantly smaller (and therefore more realistic) than the value in the corona: for the weakest magnetic fields considered the midplane $\alpha$ becomes as low as $7\times10^{-2}$, though this is still a bit higher than the value generally obtained in numerical simulations.

 \begin{figure*} 
   \centering
   \includegraphics[width= 12cm]{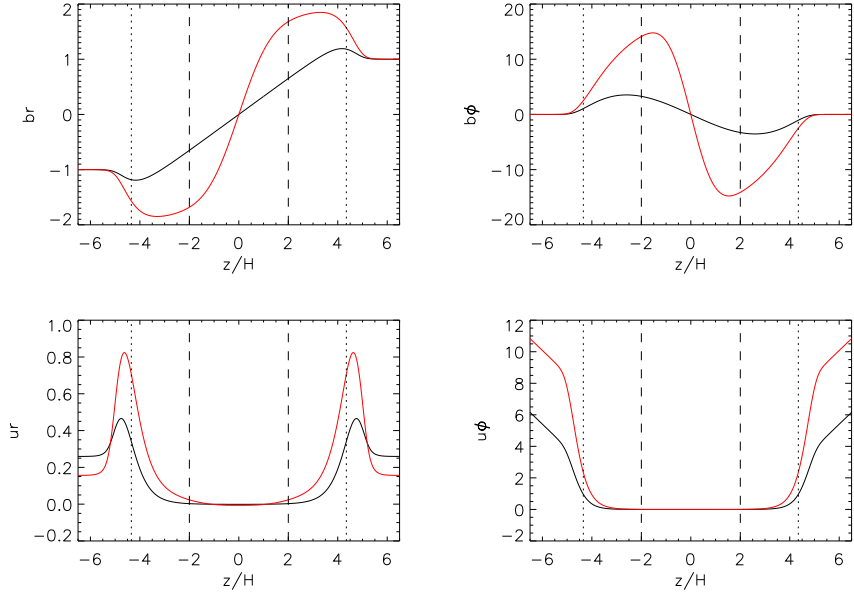}
   \caption{Diffusion of a radially inclined magnetic field with a vertical profile of the viscosity corresponding to a uniform stress in the main body of the disc. The four panels correspond to the vertical profiles of the radial (left) and azimuthal (right) magnetic field (top) and velocity (bottom) when a radial magnetic field $b_{r\mathrm{s}}=1$ ($-1$) is imposed at the upper (lower) boundary (i.e. the solution vector  $\bmath{X}_{br\rms}$). The red line corresponds to the diffusion coefficient profile of Section~\ref{sec:cst_stress} (equation~\ref{eq:exp_alpha}), while the black line is the uniform diffusion coefficient case studied in Paper~I. The strength of the vertical magnetic field is $\beta_0 =10^4$. The vertical dashed and dotted lines have the same meaning as in Figure~\ref{fig:shape_alpha_exp}.}
   \label{fig:expalpha_brs0}
 \end{figure*}

Figure~\ref{fig:expalpha_brs0} shows the vertical profiles of the velocity and magnetic field corresponding to the source term $b_{r\rms}$, which describes the vertical diffusion of the radial component of the magnetic field through the disc. The solution obtained with the diffusion-coefficient profiles given by equation~(\ref{eq:exp_alpha}) (shown with red lines) is compared to the solution obtained with uniform diffusion coefficients (shown with black lines). With uniform diffusion coefficients, the radial component of the magnetic field has a linear dependence on $\zeta$ for $|\zeta|<\zeta_B$. The shape is more complicated with the new $\alpha$ profile: the bending of the field line occurs preferentially close to the midplane ($|\zeta|<\zeta_c$) where $\alpha$ is low. This can be understood using the fact that the magnetic flux transport velocity $u_\psi$ expressed by equation~(\ref{upsi}) is independent of the vertical coordinate. Indeed, for $|\zeta|<\zeta_B$ the magnetic field behaves approximately passively, therefore the radial velocity vanishes. The derivative of the radial magnetic field can then be directly expressed as
\begin{equation}
\p_\zeta b_r =  \frac{\Pm}{\alpha} u_\psi + D_B,
	\label{eq:shape_dzbr}
\end{equation}
which shows that the bending occurs preferentially where the resistivity $\alpha/\Pm$ is low. Note that this property is the physical reason behind the definition of the average resistivity $\bar{\eta}$ to be the inverse of the average conductivity: more weight is given to the region where the conductivity is high because the magnetic field line bends preferentially there. 

Another difference between the two diffusion coefficient profiles is the amplitude of the jump in radial magnetic field around $\zeta_B$: the jump is more pronounced with the new profile with $\Delta b_r \simeq 1$, while it is $\Delta b_r \simeq 0.5$ for uniform diffusion coefficients (see Paper~I). This is intimately related to the larger bump in radial velocity around $\zeta_B$ due to the requirement that magnetic flux transport velocity be independent of $\zeta$ (equation~\ref{upsi}). A description of the transition region (between the passive magnetic field regime at $|\zeta|<\zeta_B$ and the force-free magnetic field regime at $|\zeta|>\zeta_B$) was given in Appendices~A and~B of Paper~I. It shows that the jump in $b_r$ around $\zeta_B$ is related to the jump in azimuthal velocity through $\Delta b_r = - \frac{2}{3}\frac{\Delta u_\phi}{\zeta_B}$. The numerical solutions show that the azimuthal velocity jump is approximately twice larger with the new viscosity profile, which is in agreement with this relationship. The jump in azimuthal velocity is in turn determined by the requirement that the radial electric current be continuous across $\zeta_B$: $\Delta u_\phi = (\alpha/\Pm)\p_\zeta b_\phi(\zeta_B^-)$ (equation~100 of Paper~I). The larger jumps of $b_r$ and $u_\phi$ in the new profiles can then be traced back to the fact that the azimuthal magnetic field is larger because the smaller resistivity near the midplane allows more shearing of the radial field into an azimuthal field.   

 \begin{figure*} 
   \centering
   \includegraphics[width= 12cm]{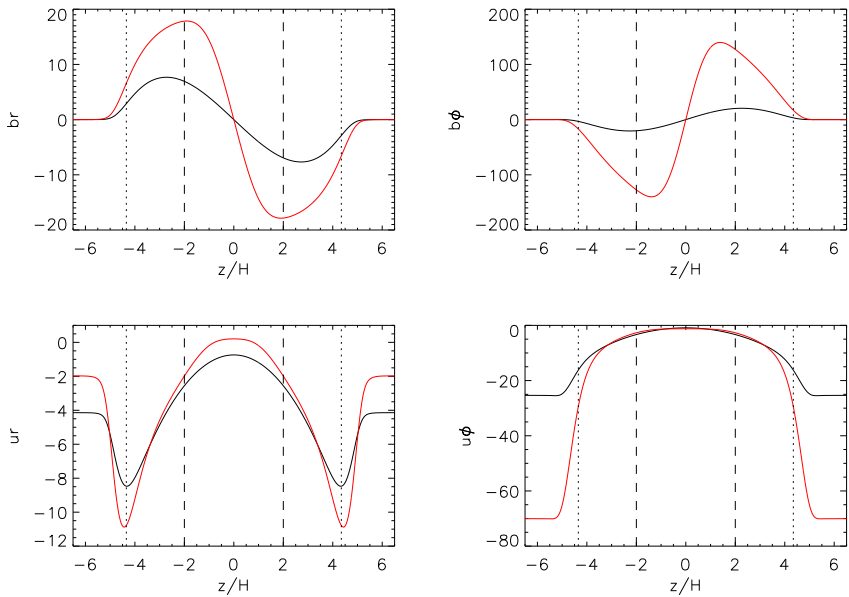}
   \caption{Same as Figure~\ref{fig:expalpha_brs0} but for the `hydrodynamic' source terms ($D_H=1$ and $D_{\nu\Sigma}=0$).}
   \label{fig:expalpha_hydro0}
 \end{figure*}

Figure~\ref{fig:expalpha_hydro0} shows the vertical profiles of the velocity and magnetic field for the hydrodynamical source terms, describing the advection due to angular momentum transport by the effective viscosity. At $|\zeta|<\zeta_c$, the radial velocity is more negative in the case of uniform diffusion coefficients, which can be interpreted as due to the larger viscosity there compared to the more realistic profile. Note that for the latter profile, the radial velocity is even positive at $|\zeta| < 0.78$ with a small maximum value of $u_r=0.2$. Such a positive radial velocity near the midplane of an accretion disc where the averaged radial velocity is negative is sometimes called meridional circulation. It has already been found in viscous models of accretion discs with uniform $\alpha$ for example by \citet{kley92} and \citet{takeuchi02}. Such a phenomenon is absent from the uniform $\alpha$ case because of the large value of $\alpha$ (equation~87 of Paper~I shows that it would appear at small $\alpha$). At $|\zeta|>\zeta_c$, the radial velocities of the two viscosity profiles become comparable as the viscosity also becomes comparable (see Figure~\ref{fig:shape_alpha_exp}). As with the source term $b_{r\rms}$, the smaller resistivity near the midplane leads to an easier bending of the field lines and therefore larger values of the radial and azimuthal magnetic field components (and therefore also a larger jump of the azimuthal velocity around $\zeta_B$).

 \begin{figure*} 
   \centering
   \includegraphics[width= 12cm]{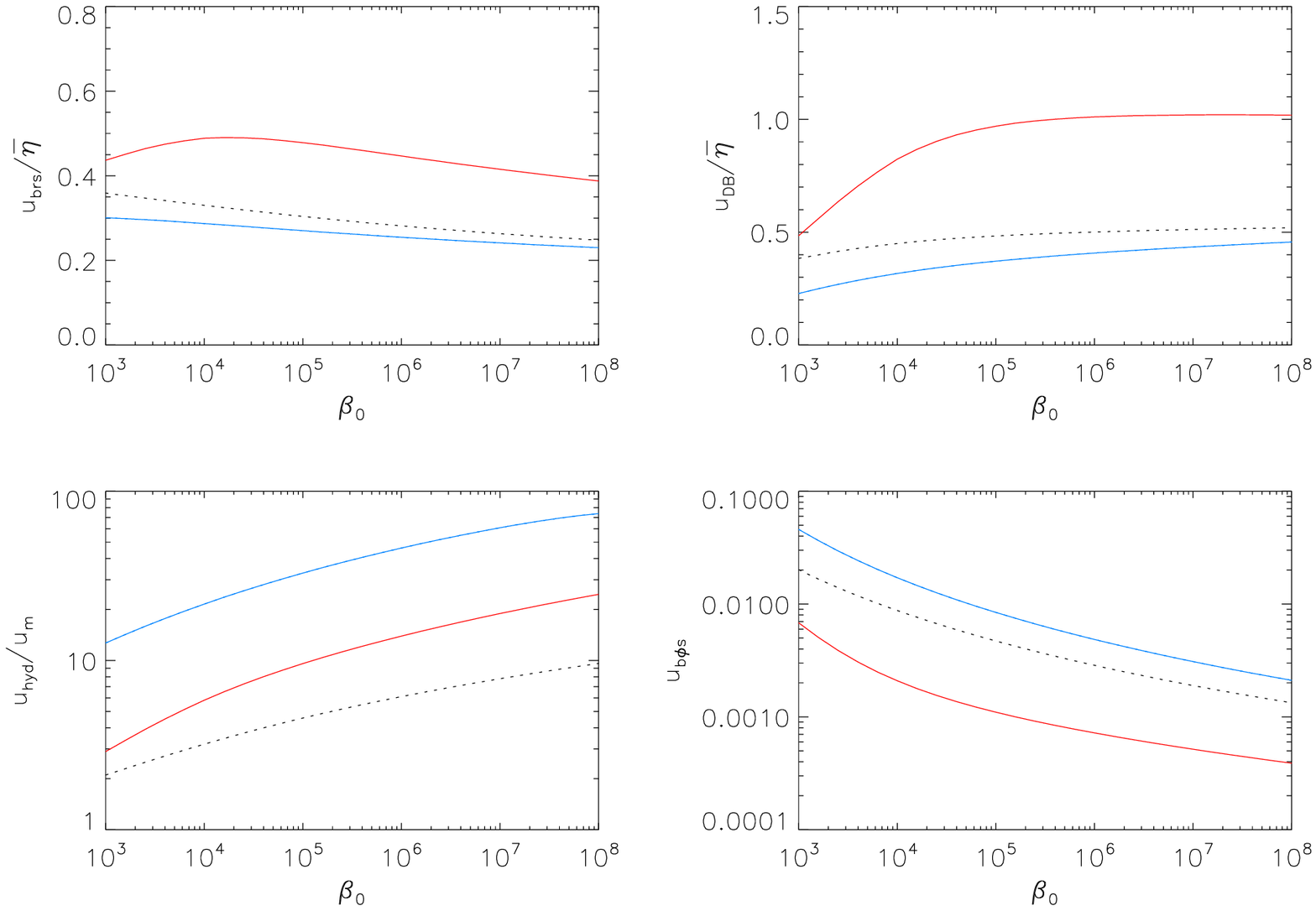}
   \caption{Contributions to the transport velocity of the magnetic flux, $u_\psi$, as a function of the magnetic field strength, for different source terms. Upper left panel: diffusion due to the bending across the disc ($b_{r\rms}$ source term). Upper right panel: diffusion due to the radial
     gradient of $B_z$ ($D_B$). These two transport velocities are normalised by the average resistivity defined in equation~(\ref{eq:eta_average}). Lower left panel : advection due to the turbulent viscosity (`hydrodynamic' source terms, $D_H=1$ and $D_{\nu\Sigma}=0$) normalised by the advection velocity of mass. Lower right panel: advection due to angular momentum removal by an outflow ($b_{\phi\rms}$). In all panels, the dotted black line corresponds to the case of uniform diffusion coefficients studied in Paper~I, the red line corresponds to the viscosity profile giving a uniform stress in the main body of the disc (Section~\ref{sec:cst_stress}) and the blue line corresponds to diffusion-coefficient profiles modelling a dead zone (Section~\ref{sec:dead_zone}). }
   \label{fig:deadexp_uflux}
 \end{figure*}
 
 \begin{figure*}
   \centering
   \includegraphics[width=2\columnwidth]{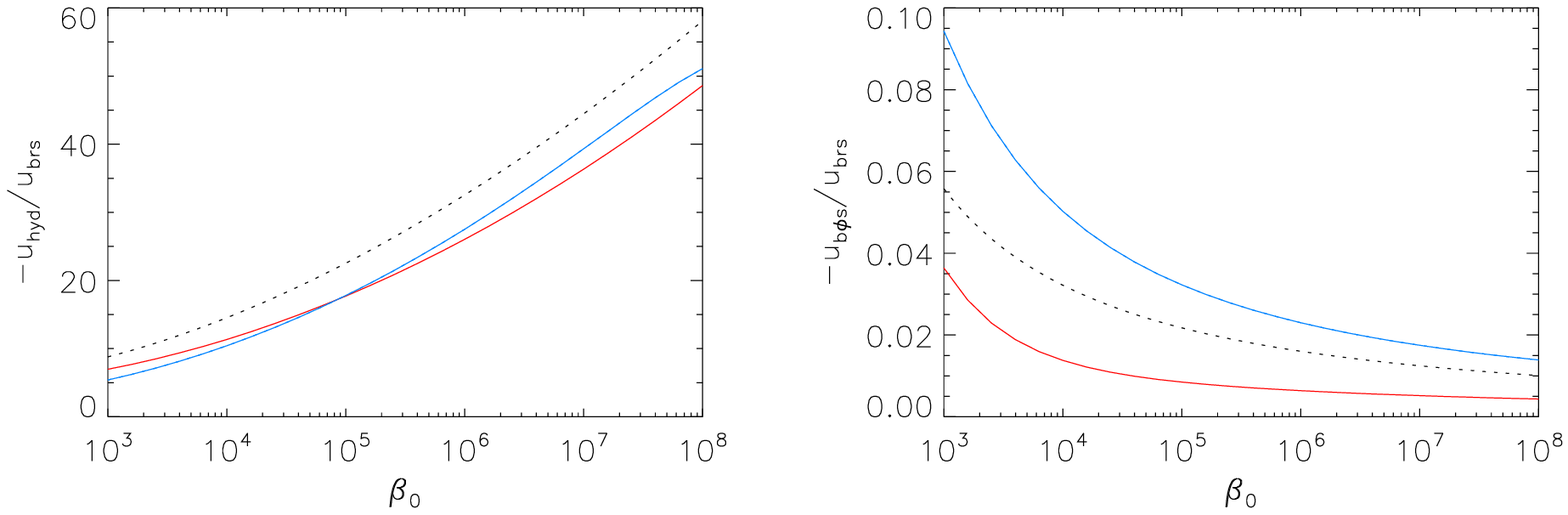}
   \caption{Radial inclination of the field lines at the surface of the disc obtained in a stationary situation. The left panel shows the inclination ($B_{r\rms}/B_z$) in units of $H/r$ such that the diffusion compensates the advection due to the turbulent viscosity (`hydrodynamic' source terms). The right panel shows the ratio $-b_{r\rms}/b_{\phi\rms}$ such that diffusion compensates advection due to an outflow. In both panels, the dotted black line corresponds to the case of uniform diffusion coefficients studied in Paper~I, the full red line corresponds to the profile with a constant stress in the midplane (Section~\ref{sec:cst_stress}), and the full blue line corresponds to the dead-zone model (Section~\ref{sec:dead_zone}).}
   \label{fig:deadexp_inclination}
\end{figure*}

Figure~\ref{fig:deadexp_uflux} shows the transport velocity of the magnetic flux for the different source terms (red lines) and compares it to the case of a uniform
$\alpha$ (black lines). The transport velocity due to the inclination of the field lines ($b_{r\rms}$) is normalised by the averaged resistivity $\bar{\eta}$ as suggested by the averaging procedure described in Section~\ref{sec:induction_average} (it is shown in the upper left panel). Normalised in this way, the diffusion velocity is a bit larger with the new profile of $\alpha$ because of the larger jump in $b_r$ discussed above. All in all this diffusion velocity is, however, smaller with the new profile due to the lower value of $\bar{\eta}$, as can be seen on Figure~\ref{fig:expalpha_brs0} by comparing the radial velocity at large $\zeta$. The diffusion velocity associated with the source term $D_B$ (radial gradient of magnetic field strength), normalised in the same way, is also larger with the new $\alpha$ profile (upper right panel). This is due to the same phenomenon: the jump in $b_r$ is larger because the resistivity is reduced in the midplane\footnote{The difference in diffusion velocity between the two $\alpha$ profiles is more pronounced for the source term $D_B$ than for the source term $b_{r\rms}$. This is because all of the diffusion is due to the jump in $b_r$ in the case of $D_B$, while part or most of it is not due to this jump in the case of $b_{r\rms}$.}. The transport velocity of the magnetic flux due the hydrodynamical source term, shown in the lower left panel, is normalised by the transport velocity of mass. These two transport velocities were assumed to be the same in \citet{lubow94a} and other papers, but we showed in Paper~I that they can differ significantly by a factor of up to 10 for the magnetic field strength considered. This is explained by the fact that the transport velocity of mass is dominated by the region close to the midplane (since it is a density-weighted average) while the transport velocity of magnetic flux is significantly affected by the low-density region away from the midplane (since it is a conductivity weighted average) where the velocity is large. The same qualitative feature is found with the more realistic viscosity profile, but the ratio $u_\psi/u_m$ is even higher, reaching up to 25 for $\beta_0=10^8$. This is a consequence of the larger difference between the (smaller) midplane velocity and the (similar) velocity in the corona. Finally the transport velocity of magnetic flux due to an outflow (source term $b_{\phi\rms}$) is shown in the lower right panel. It is found to be smaller with the new diffusion coefficient profile by a factor of $\simeq 3$.  

An important consequence of the relative efficiency of the diffusion and advection processes is the radial inclination with respect to the vertical direction that magnetic field lines can sustain in a stationary situation. When the advection process is due to the angular momentum transport by the effective viscosity, the inclination is expressed as (see Section~4.6 of Paper~I)
\begin{equation}
\frac{B_{r\rms}}{B_z} = -\frac{H}{r}\frac{u_\mathrm{hyd}}{u_{br\rms}}.
	\label{eq:inc_hydro}
\end{equation}
This inclination in units of $H/r$ is shown in the left panel of Figure~\ref{fig:deadexp_inclination}. Interestingly this inclination has a very similar behaviour with the new diffusion-coefficient profile as with uniform diffusion coefficients (the two differ by about $20\%$): it increases as the magnetic field strength is decreased, and reaches quite large values of up to $50$ (to be compared to $3/2$ used by \citealt{lubow94a}). This shows that this important result of Paper~I is robust and remains true for more realistic treatment of the vertical structure of the turbulence. 

If the advection of magnetic flux is due to to angular momentum loss in an outflow, the ratio of radial to azimuthal magnetic field at the surface of the disc in a stationary situation is determined by
\begin{equation}
\frac{B_{r\rms}}{B_{\phi\rms}} = - \frac{u_{b\phi\rms}}{u_{br\rms}}.
	\label{eq:inc_bphis}
\end{equation}
This quantity is shown in the right panel of Figure~\ref{fig:deadexp_inclination}. It is smaller with the new profile because of the lesser efficiency with which the accretion flow driven by an outflow can transport magnetic flux.

\section{Disc containing a dead zone}
\label{sec:dead_zone}

\begin{figure*}
   \centering
   \includegraphics[width=\columnwidth]{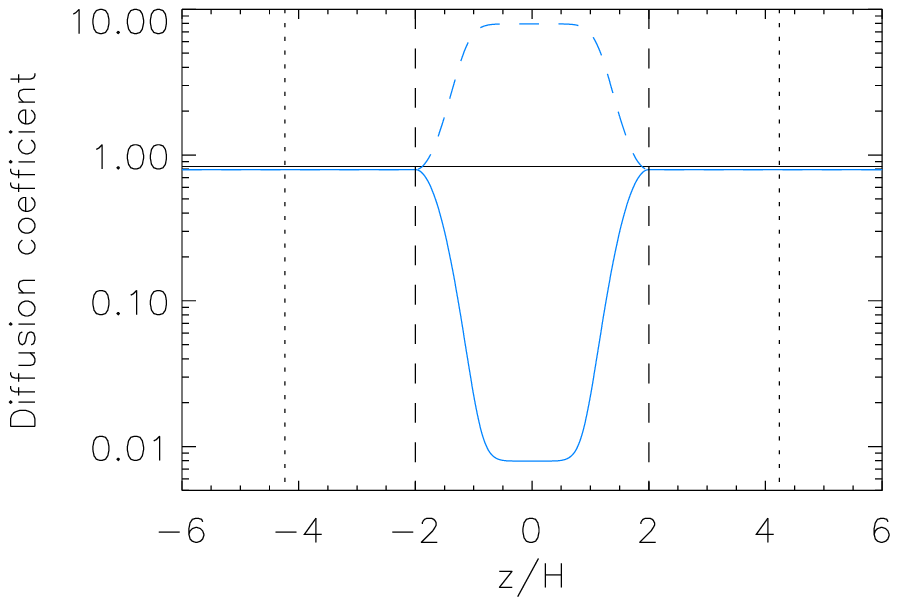}
    \includegraphics[width=\columnwidth]{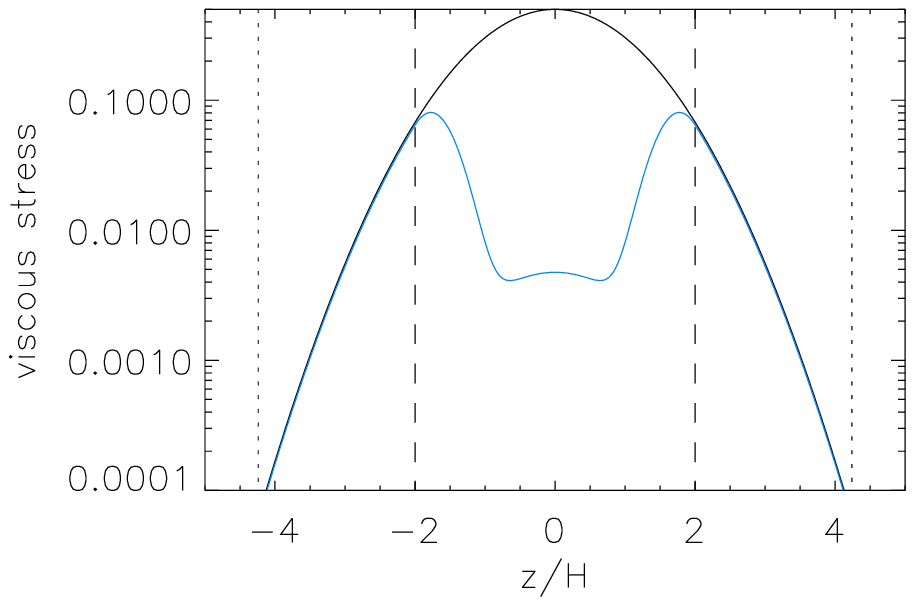}
   \caption{Disc containing a dead zone: vertical profiles of the diffusion coefficients (left panel), and of the viscous stress ($r\phi$ component of the viscous stress tensor) $(3/2)\tilde{\rho}\alpha$ (right panel) for a magnetic field strength $\beta_0=10^4$. In the left panel, the full line shows the dimensionless viscosity $\alpha$, and the dashed line the dimensionless resistivity $\alpha/\Pm$. In both panels, the blue lines show the profiles used in Section~\ref{sec:dead_zone} to model a dead zone (equations~(\ref{eq:dead_alpha0})-(\ref{eq:dead_eta})), while the black line shows uniform diffusion coefficient case studied in Paper~I. The vertical dashed lines show the height $\zeta_\mathrm{cut}$ delimiting the dead zone, and the dotted line the height $\zeta_B$ above which the magnetic field is dominant. }
   \label{fig:shape_alpha_dead}
\end{figure*}

\begin{figure}
   \centering
   \includegraphics[width=\columnwidth]{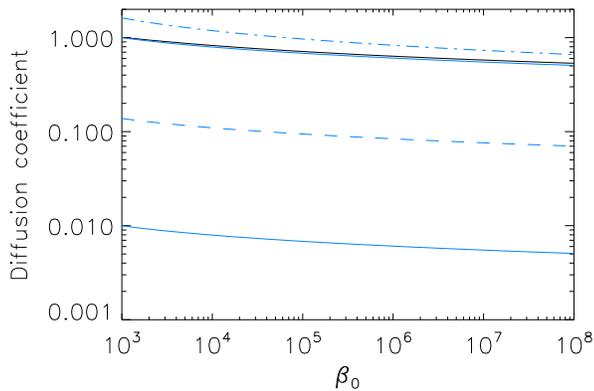}
   \caption{Diffusion coefficients determined by the marginal stability hypothesis as a function of the magnetic field strength for the dead zone model of Section~\ref{sec:dead_zone}. The black line shows the uniform $\alpha$ case for comparison. Blue lines correspond to the profile modelling a dead zone given in equations~(\ref{eq:dead_alpha0})-(\ref{eq:dead_eta}): the minimum and maximum values of $\alpha$ are represented by full lines, the density-weighted average by the dashed line, and the average resistivity (Equation~\ref{eq:eta_average}) by the dash-dotted line.}
   \label{fig:alpha_beta_dead}
\end{figure}

The diffusion-coefficient profile studied in Section~\ref{sec:cst_stress} models an accretion disc where MHD turbulence operates throughout the disc. This is expected for discs which are sufficiently ionised to enable the MRI to operate, such as the hot discs around black holes or a neutron star in AGN and X-ray binaries. The cooler discs around a T~Tauri star may not be sufficiently ionised throughout their volume to sustain MHD turbulence, and a laminar region called a `dead zone' may then exist \citep{gammie96}. Determining the transport of magnetic flux in such discs then requires a specific modelling of this dead zone, which is the purpose of this section. Understanding the strength of the magnetic field in protoplanetary discs is important for the launching of outflows as outlined in the introduction \citep{ferreira06}, but also for the theory of planet formation. Indeed dead zones are a probable formation site for planets, and recent studies show that the magnetic field could strongly affect the migration of planets \citep[Guilet et al., submitted to MNRAS]{terquem03,fromang05}.

A disc can sustain MHD turbulence if it has a sufficiently low physical resistivity, which is governed mostly by the ionisation fraction of the gas. Different physical processes are responsible for the ionisation depending on the region of the disc considered. In the inner parts of protoplanetary discs ($r \lesssim 1\, {\rm AU}$) the temperature can be high enough for the thermal ionisation of metals to provide sufficient ionisation (if $T\gtrsim 1000\, {\rm K}$). At larger radii, non-thermal processes are the main agent of ionisation: X-rays or ultraviolet light emitted by the central star or possibly cosmic rays. These ionising radiations are able to penetrate only a finite surface density into the disc (of the order $100\, \mathrm{g}\,\mathrm{cm}^{-2}$, though this depends on the energy of the photons or cosmic rays). As a consequence, they can ionise the whole vertical extent of the disc only at large disc radii, typically $r\gtrsim 5-10\, {\rm AU}$. At intermediate radii, only the surface layer of the disc is sufficiently ionised to sustain MHD turbulence, while a region around the midplane has a too high resistivity and therefore remains laminar (the dead zone). Note that the ionisation fraction is very sensitive to the amount of small dust particles and metal in the gaseous phase, and as a consequence is very model-dependent \citep{fromang02}. The extent of the dead zone is therefore also very uncertain and the numbers quoted above should be considered as only rough estimates.    

In order to model the vertical structure of the region of a protoplanetary disc that contains a dead zone, we choose the following functional form of the diffusion coefficients:
\begin{eqnarray}
\alpha(|\zeta| > \zeta_\mathrm{cut}) &=& \alpha_0, \label{eq:dead_alpha0} \\
\alpha/\Pm(|\zeta| > \zeta_\mathrm{cut}) &=& \alpha_0/\Pm_0, \\
\alpha(|\zeta| < \zeta_\mathrm{cut}) &=& \alpha_0 \left\lbrack \nu_\mathrm{cut} + (1-\nu_\mathrm{cut})e^{-(|\zeta|-\zeta_\mathrm{cut})^2/L_\mathrm{cut}^2}\right\rbrack, \label{eq:dead_alpha} \\
\alpha/\Pm(|\zeta| < \zeta_\mathrm{cut}) &=& \frac{\alpha_0/\Pm_0} {\left\lbrack \eta_\mathrm{cut} + (1-\eta_\mathrm{cut})e^{-(|\zeta|-\zeta_\mathrm{cut})^2/L_\mathrm{cut}^2}\right\rbrack}, \label{eq:dead_eta}
\end{eqnarray}
where $\zeta_\mathrm{cut}$ is the height up to which the dead zone extends, $\nu_\mathrm{cut}$ is the factor by which the viscosity is reduced inside the dead zone, $1/\eta_\mathrm{cut}$ the factor by which the resistivity is increased, and $L_\mathrm{cut}$ is the width of the transition between the active layer and the dead zone. As an illustrative example, we choose the following values: $\zeta_\mathrm{cut}=2$, $\nu_\mathrm{cut}=10^{-2}$, $\eta_\mathrm{cut}=10^{-1}$ and $L_\mathrm{cut}=0.5$. The viscosity and resistivity profiles are shown in the left panel of Figure~\ref{fig:shape_alpha_dead} for a magnetic field strength of $\beta_0=10^4$. In the active layers above $\zeta_\mathrm{cut}$, the viscosity and resistivity are very close to the case of uniform diffusion coefficients. Inside the dead zone the resistivity increases by a factor $10$ while the viscosity is decreased by a factor of $100$. The corresponding viscous stress profile is shown in the right panel of Figure~\ref{fig:shape_alpha_dead}. Inside the dead zone it decreases to a minimum value which is about $10\%$ of the maximum stress reached in the active layer. This is consistent with the result of numerical simulations of local disc models containing a dead zone. These find that the stress in the dead zone is dominated by the reynolds stress due to waves excited at the boundary with the active layers, and is about $10\%$ of the stress in the active layers \citep{fleming03}. 

 \begin{figure*}
   \centering
   \includegraphics[width= 12cm]{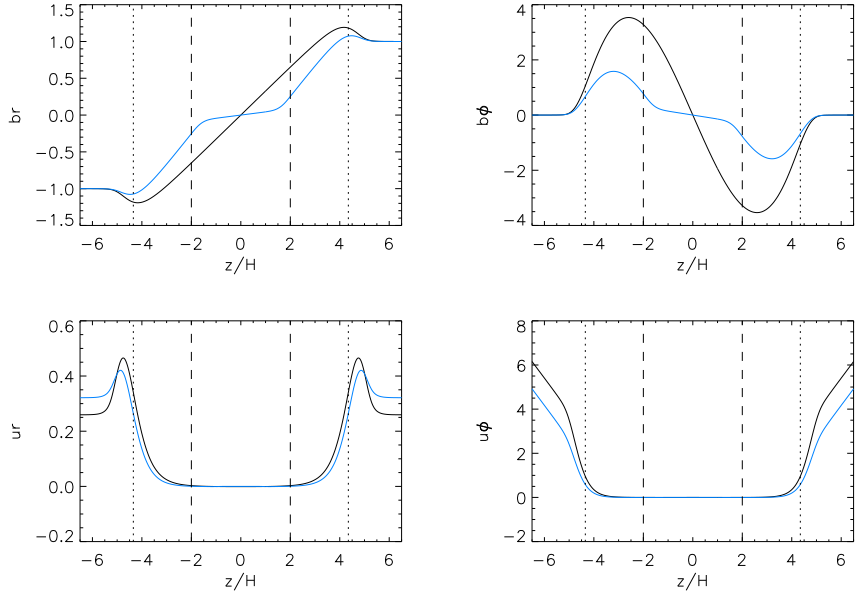}
   \caption{Same as Figure~\ref{fig:expalpha_brs0} but with diffusion coefficients profiles modelling a dead zone (shown with blue lines). The vertical dashed lines show the vertical extent of the dead zone.}
   \label{fig:dead_zone_brs}
 \end{figure*}

 \begin{figure*}
   \centering
   \includegraphics[width= 12cm]{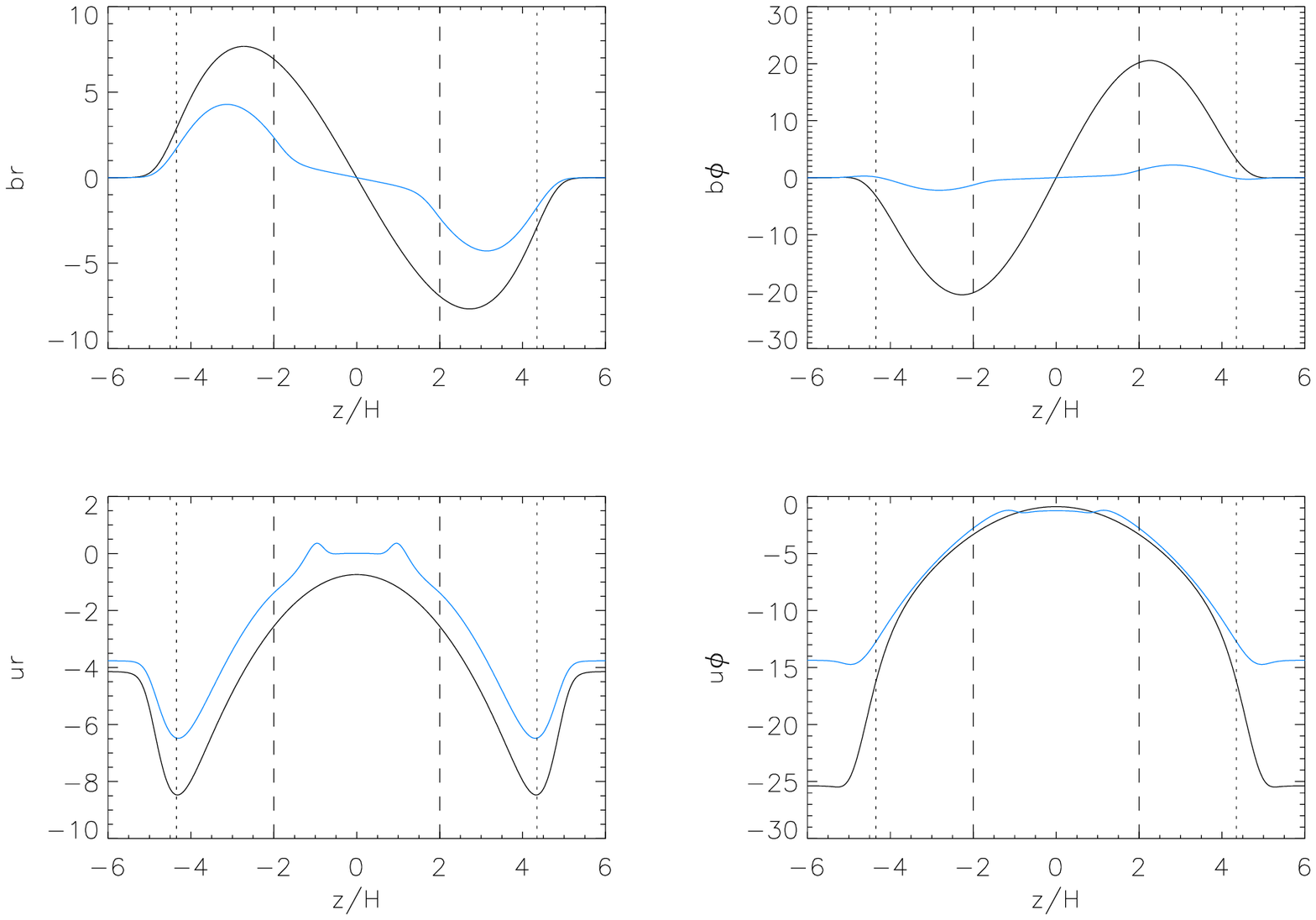}
   \caption{Same as Figure~\ref{fig:expalpha_hydro0} but with diffusion coefficients profiles modelling a dead zone (shown with blue lines). The vertical dashed lines show the vertical extent of the dead zone.}
   \label{fig:dead_zone_hydro}
 \end{figure*}

The dependence with respect to the magnetic field strength of various measures of the diffusion is shown in Figure~\ref{fig:alpha_beta_dead}. The density-weighted averaged viscosity (relevant to the transport of mass) is reduced by a factor $\simeq 7$ due to the low viscosity in the dead zone. The averaged resistivity is only slightly larger than the resistivity in the active layers, although it is much larger in the dead zone. This is a consequence of the averaging procedure used: the volume average of the resistivity would be much larger than this inverse of the averaged conductivity.   
 
The vertical profiles of the velocity and the magnetic field obtained for the source term $b_{r\rms}$, describing the vertical diffusion of a radial magnetic field, are shown in Figure~\ref{fig:dead_zone_brs}. The dead zone model (blue lines) is compared to the uniform diffusion coefficient profile of Paper~I (black lines). The main difference is that the large resistivity inside the dead zone prevents the magnetic field lines from bending there, such that the radial and azimuthal components of the magnetic field remain close to zero for $\zeta<\zeta_\mathrm{cut}$. All the bending therefore occurs in the active layers at $\zeta_\mathrm{cut}<\zeta<\zeta_B$, where the profiles are qualitatively similar to the uniform diffusion coefficient case; in particular the radial component of the magnetic field has a linear dependence on $\zeta$ because the diffusion coefficients are uniform in the active layers. The transport velocity of the magnetic flux (visualised as the radial velocity at large $|\zeta|$) is slightly larger in the presence of the dead zone, consistent with the slightly larger average resistivity. This diffusion velocity normalised by the average resistivity is shown as a function of the magnetic field strength in the upper left panel of Figure~\ref{fig:deadexp_uflux}. It is slightly smaller in the dead zone model than in the uniform diffusion coefficient case, due to the fact that the bump in $b_r$ around $\zeta_B$ is slightly less pronounced in the presence of a dead zone.

The vertical profiles of the velocity and magnetic field for the hydrodynamical source terms are shown in Figure~\ref{fig:dead_zone_hydro}. Again the radial and azimuthal magnetic field components remain close to zero inside the dead zone due to the large resistivity there. The radial velocity in the dead zone is slightly positive for $|\zeta|<1.2$, and becomes comparable to the uniform diffusion coefficient case (though slightly less negative) outside of the dead zone. The advection velocity of the magnetic flux of the two models is similar. This can be interpreted by the fact that the radial velocity in the dead zone (where the two models differ most) barely affects the conductivity weighted averaged radial velocity because the conductivity is very small. The ratio of the magnetic flux and mass transport velocities is shown in the lower left panel of Figure~\ref{fig:deadexp_uflux}: it is a factor $\simeq 7$ larger in the presence of a dead zone, because the mass transport velocity is 7 times smaller than in the case of uniform diffusion coefficients while the magnetic flux transport velocity is similar in both models.  

The left panel of Figure~\ref{fig:deadexp_inclination} shows the radial inclination of the magnetic field lines at the surface of the disc in a stationary situation where the diffusion compensates exactly the advection. It is very similar in the dead zone model and in the uniform diffusion coefficient model. Indeed we have seen that the advection velocity is very similar, and that the diffusion velocity is only very slightly larger. 

The transport velocity of magnetic flux due to an outflow is shown in the lower left panel of Figure~\ref{fig:deadexp_uflux}. It is a factor $\simeq 1.5-2$ larger in the presence of a dead zone than with uniform diffusion coefficients. As a consequence the ratio of radial to azimuthal magnetic field when this transport compensates diffusion is also a factor $\simeq 1.5-2$ larger (left panel of Figure~\ref{fig:deadexp_inclination}).

\section{Non-turbulent surface layer}
	\label{sec:lovelace}

 \begin{figure*} 
   \centering
   \includegraphics[width= \columnwidth]{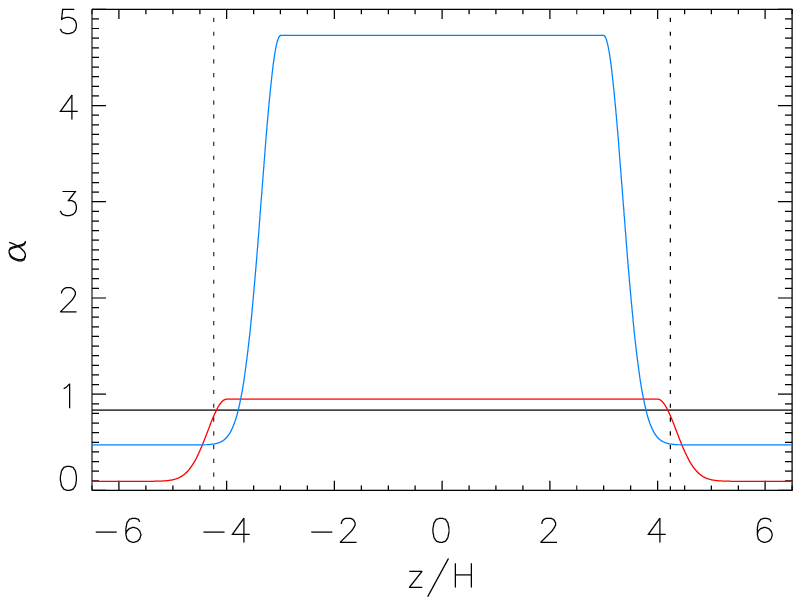} 
    \includegraphics[width= \columnwidth]{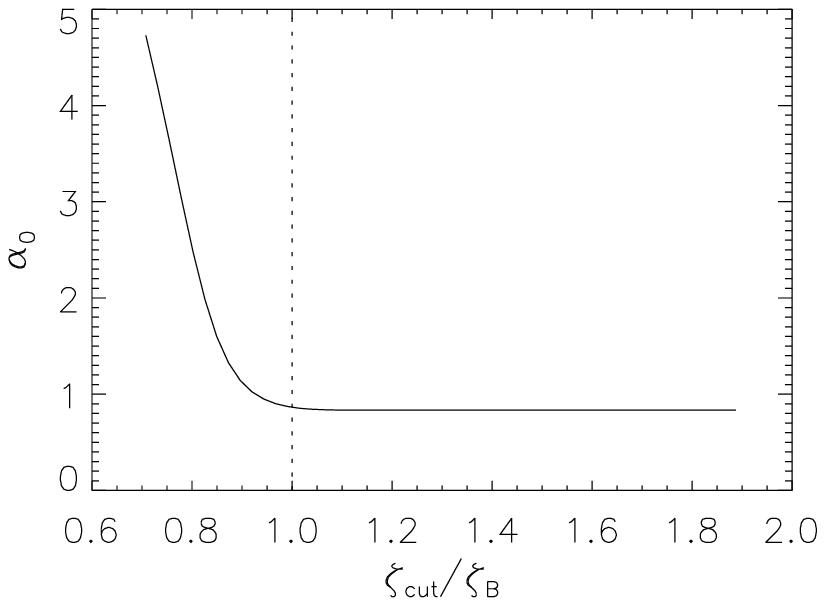}
   \caption{Diffusion coefficients in the model of a non-turbulent surface layer. Left panel: vertical profiles of the dimensionless viscosity parameter $\alpha$ for $\beta_0=10^4$. The red line corresponds to a position of the non-turbulent layer of $\zeta_\mathrm{cut}=4$, the blue line to $\zeta_\mathrm{cut}=3$, while the black line shows the uniform diffusion case of Paper~I. The vertical dotted lines show the height $\zeta_B$ above which the magnetic field is dominant. Right panel: dimensionless viscosity in the midplane $\alpha_0$ as a function of the height $\zeta_\mathrm{cut}$ of the non-turbulent surface layer normalised by the height $\zeta_B$ at which the magnetic field becomes dominant.}
   \label{fig:shapealpha_cut}
 \end{figure*}
 
In a series of four papers   \citet{bisnovatyi-kogan07}, \citet{rothstein08}, \citet{lovelace09}, and \citet{bisnovatyi-kogan12} have proposed a model in which the diffusion of the magnetic field is reduced owing to the existence of a non-turbulent layer near the surface of the accretion disc. In this model all the bending of the field lines takes place in this surface layer without causing any diffusion because the conductivity is very high (due to the lack of turbulence). This model however still lacks a proper calculation of the vertical profile of the magnetic field showing such an effect. Indeed the first two papers use a heuristic approach and do not attempt such a calculation. The last two papers \citep{lovelace09,bisnovatyi-kogan12} describe a formalism that in principle takes into account a non-turbulent surface layer. But solutions are then only calculated in the particular case where the diffusivity and density are uniform inside the disc, and where the diffusivity vanishes only in the wind outside the disc, which is not computed explicitly. As a result there is no change of the radial magnetic field in the surface layer, and therefore the diffusive transport of the magnetic flux is not decreased. In these papers the main ingredient allowing a significant bending of the magnetic field lines inside the disc is an increased advection due to the presence of an outflow, rather than a decreased diffusion due to a non-turbulent surface layer. This possibility was discussed in paper~I. The purpose of this section is to test the idea proposed in \citet{bisnovatyi-kogan07} and \citet{rothstein08} that a non-turbulent surface layer can reduce the diffusion of the magnetic flux. To this end, we perform a calculation of the vertical profiles of velocity and magnetic field in a disc with a non-turbulent surface layer and determine the associated diffusion.

In order to model the non-turbulent surface layer, we choose the following vertical profile of $\alpha$:
\begin{eqnarray}
\alpha(|\zeta| < \zeta_\mathrm{cut}) &=& \alpha_0, \\
\alpha(|\zeta| > \zeta_\mathrm{cut}) &=& \alpha_0 \left\lbrack
\alpha_\mathrm{cut} + (1-\alpha_\mathrm{cut})e^{-(|\zeta|-\zeta_\mathrm{cut})^2/L_\mathrm{cut}^2}\right\rbrack,
\label{eq:cut_alpha}
\end{eqnarray}
where $\zeta_\mathrm{cut}$ is the height above which the viscosity and resistivity are reduced (and below which they are uniform). Above $\zeta_\mathrm{cut}$, $\alpha$ transitions
from $\alpha_0$ to $\alpha_\mathrm{cut}\alpha_0$ over a height of $L_\mathrm{cut}$. The magnetic Prandtl number $\Pm$ is assumed to be uniform and equal to $1$. In the following we used the values $L_\mathrm{cut}=0.5$, and $\alpha_\mathrm{cut}=0.1$ (we checked that decreasing this factor further does not change qualitatively the results). The location of the transition to a non-turbulent state $\zeta_\mathrm{cut}$ was varied to determine where it should lie for the non-turbulent layer to effectively reduce the diffusion of magnetic flux. As discussed by \citet{rothstein08} the physically motivated position of the non-turbulent layer is $\zeta_\mathrm{cut} \simeq \zeta_B$, where the magnetic pressure equals the thermal pressure. Indeed, above $\zeta_B$ the magnetic pressure exceeds the thermal pressure and as a consequence the MRI is stabilised, resulting in a laminar state, although the reality may be more complicated.  

The left panel of Figure~\ref{fig:shapealpha_cut} shows the vertical profiles of the diffusion coefficients for $\beta_0=10^4$ and for two positions of the non-turbulent surface layer: $\zeta_\mathrm{cut}=3$ (blue line) and $\zeta_\mathrm{cut}=4$ (red line) which is close to the expected value of $\zeta_B\simeq4.3$. These two profiles are compared to the uniform diffusion coefficient case shown with a black line. For $\zeta_\mathrm{cut}=4$, the marginal stability analysis gives a very similar diffusion coefficients value inside the disc for both of the uniform coefficient case and the non-turbulent layer model. This is consistent with the idea that above $\zeta_B$ the MRI is stabilised because of the strength of the magnetic field: no diffusion is therefore needed in this region to achieve marginal stability. For $\zeta_\mathrm{cut}=3$ ($<\zeta_B$), the marginal stability analysis gives much larger diffusion coefficients inside the disc. This shows that for this small value of $\zeta_\mathrm{cut}<\zeta_B$, the magnetic field in the region of decreased diffusion is not large enough to stabilise the MRI, such that the diffusion has to be increased to achieve marginal stability. This idea is further demonstrated in the right panel of Figure~\ref{fig:shapealpha_cut}, showing $\alpha_0$ (the value of $\alpha$ inside the disc) as a function of the position of the non-turbulent layer $\zeta_\mathrm{cut}$ normalised by its physically expected value $\zeta_B$. For $\zeta_\mathrm{cut}>\zeta_B$, $\alpha_0$ is unchanged by the inclusion of the non-turbulent layer. On the contrary, for $\zeta_\mathrm{cut}<\zeta_B$, $\alpha_0$ increases strongly when $\zeta_\mathrm{cut}$ decreases, showing that we are trying to decrease the diffusion in a region where the MRI is {\it not} stable. As a conclusion, the marginal stability analysis confirms that the expected position of the non-turbulent layer is $\zeta_\mathrm{cut}\simeq\zeta_B$.


 \begin{figure*}
   \centering
   \includegraphics[width= 12cm]{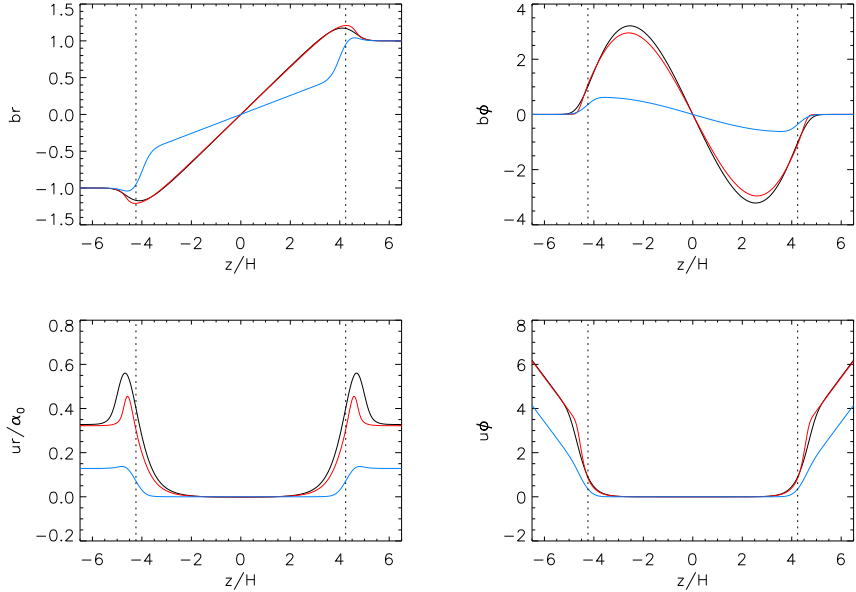}
   \caption{Magnetic field diffusion in the presence of a non-turbulent surface layer: vertical profiles of the radial (left) and azimuthal (right) velocity (upper panels) and magnetic field (lower panels) for the source term $b_{r\rms}$. Note that the radial velocity is normalised by $\alpha_0$, the diffusion coefficient inside the disc. The calculation using the physically motivated position of the non-turbulent layer $\zeta_\mathrm{cut}=4$ is shown with red lines, $\zeta_\mathrm{cut}=3$ is shown with a blue line and the uniform diffusion coefficient case is shown in black. The vertical dotted lines show the height $\zeta_B$ above which the magnetic pressure dominates over the thermal pressure. The magnetic field strength is $\beta_0=10^4$. The non-turbulent surface layer at its expected position barely changes the vertical profiles and the diffusion velocity. A significant bending of the field line inside the layer (as assumed in the model of \citet{bisnovatyi-kogan07}) and associated reduction of the diffusion occurs only if the non-turbulent layer is artificially extended below its expected height $\zeta_B$.}
   \label{fig:cutalpha_brs0}
 \end{figure*}

 \begin{figure}
   \centering
   \includegraphics[width=\columnwidth]{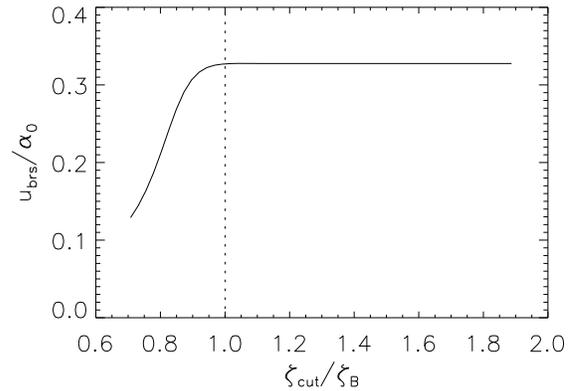}
   \caption{Diffusion of an inclined magnetic field in the presence of a non-turbulent surface layer: transport velocity normalised by $\alpha$ in the main body of the disc as a function of the height at which the non turbulent layer starts normalised by $\zeta_B$ (the height above which the magnetic field is dominant). The non-turbulent surface layer decreases the diffusion only if it extends below $\zeta_B$, which is not expected to be true.}
   \label{fig:cutalpha_uflux}
 \end{figure}

Figure~\ref{fig:cutalpha_brs0} shows the vertical profiles of the velocity and magnetic field for the source term $b_{r\rms}$, corresponding to the diffusion due to the radial inclination of the field lines at the surface of the accretion disc. The solution with the physically expected position of the non-turbulent layer is shown with red lines. It lies very close to the profiles corresponding to the uniform diffusion coefficient case (black lines). This shows that the non-turbulent surface layer at its expected position barely affects the diffusion of the magnetic field, thus contradicting the model of \citet{bisnovatyi-kogan07}. Results similar to the expectations of \citet{bisnovatyi-kogan07} are obtained only when the non-turbulent layer is artificially extended to a smaller height, as shown with a blue line for $\zeta_\mathrm{cut}=3$. In this case a significant part of the bending of the field line occurs in the non-turbulent layer (upper left panel). As a result the diffusion velocity of the magnetic flux (visible as the radial velocity at large $\zeta$) normalised by $\alpha_0$ is decreased compared to the case of uniform diffusion coefficients. This result is further demonstrated by Figure~\ref{fig:cutalpha_uflux}, showing the diffusion velocity of the magnetic flux normalised by $\alpha_0$ as a function of the height of the non-turbulent layer $\zeta_\mathrm{cut}$ normalised by its expected value $\zeta_B$. The diffusion velocity of the magnetic flux is unchanged as long as $\zeta_\mathrm{cut}>\zeta_B$. It is decreased by the presence of the non-turbulent layer only if $\zeta_\mathrm{cut}<\zeta_B$. This is consistent with the averaging procedure described in Section~\ref{sec:induction_average}: we argued that the average resistivity should be computed up to the height $\zeta_B$, therefore the non-turbulent surface layer should extend below $\zeta_B$ in order to decrease the average resistivity determining the diffusion of the magnetic flux.

The numerical results presented above disagree with the expectations of \citet{bisnovatyi-kogan07} and \citet{rothstein08}. Indeed, we find a stringent constraint on the position of the non-turbulent layer in order to decrease the magnetic flux diffusion. This constraint is intimately linked with the condition on the magnetic field strength that should be fulfilled to enable a bending of the magnetic field lines. In \citet{bisnovatyi-kogan07} and \citet{rothstein08} this condition was evaluated by comparing the radial magnetic force to the gravity. This may however be a very optimistic choice because it would correspond to a non-rotating surface layer, which seems a bit extreme. The asymptotic expansion shows that the deviation from Keplerian rotation is of the order of $\delta v_\phi \sim \frac{B_r}{B_z}c_\mathrm{s} \sim \frac{H}{r}\frac{B_r}{B_z}v_\mathrm{K}$. As a result of this smaller deviation from Keplerian rotation, the condition for a significant bending of the magnetic field lines on a length-scale of order $H$ is: $\beta \gtrsim 1$ or equivalently $\zeta\lesssim \zeta_B$. This is consistent with the constraint on the position of the non-turbulent layer $\zeta_\mathrm{cut}<\zeta_B$ so that the field lines can bend there. 

Note that with $\zeta_\mathrm{cut}=\zeta_B$, the above argument would still leave some hope that a significant bending could occur in the non-turbulent layer if this bending can occur for $\beta\sim1$. And the jump in $b_r$ observed to occur around $\zeta_B$ shows that such bending can indeed occur. However, this bending is no longer controlled only by the diffusion as in the passive field region, but also by dynamical constraints since the magnetic force is no longer negligible. As a result it is crucial to self-consistently compute the velocity and magnetic field profile to determine what bending can occur in this region. The numerical results show that the bending of the field line around $\zeta_B$ is always in the wrong direction: the radial magnetic field decreases around $\zeta_B$ after reaching a maximum at $\zeta\lesssim\zeta_B$. This feature is robust and was obtained in Paper~I as well as Section~\ref{sec:cst_stress} and \ref{sec:dead_zone}. It was interpreted in Paper~I as due to the fact that the azimuthal velocity at $\zeta\gtrsim\zeta_B$ is super-Keplerian when the magnetic field lines are bending outward ($b_{r\rms}>0$). To enable super-Keplerian velocities the magnetic force should be directed inwards and therefore the radial magnetic field should decrease with height. As a result (as shown in Figure~\ref{fig:cutalpha_brs0}) when $\zeta_\mathrm{cut}\simeq\zeta_B$, the non-turbulent surface is inefficient at reducing the diffusion.

\section{Conclusions}
	\label{sec:conclusion}
\subsection{Summary}
We have studied the dependence of the transport of magnetic flux on the vertical profile of the diffusion coefficients. For this purpose we have examined three different physically motivated vertical profiles. The first profile (Section~\ref{sec:cst_stress}) is inspired by the results of numerical simulations of MHD turbulent stratified discs. It therefore aims at describing fully turbulent discs in a more realistic way than the uniform diffusion coefficient model presented in Paper~I. We confirm the main results of Paper~I with this more realistic treatment, thus showing their robustness: the magnetic flux is advected more easily for weak magnetic field strength, resulting in a steady-state radial inclination of the magnetic field lines that can be up to 30 times larger than predicted by previous works \citep{lubow94a,heyvaerts96}.

In Section~\ref{sec:dead_zone}, we studied a model of disc containing a non-turbulent dead zone. We found that the presence of the dead zone does not change the radial inclination of the field lines in a steady state. It does change, however, the ratio of the advection velocity of magnetic flux to that of mass: the magnetic flux can be advected up to 100 times faster than mass for the dead-zone model described in this paper, and even more for more extended dead zones. This behaviour could lead to an interesting time-dependence of the magnetic field configuration, which has the ability to evolve much faster than the mass distribution. 

In Section~\ref{sec:lovelace}, we studied the effect of a non-turbulent surface layer on the diffusion of the magnetic flux. We found that it is able to reduce the diffusion of the magnetic flux only if it extends below the height $\zeta_B$ at which the magnetic pressure of the large scale vertical field equals the thermal pressure. Since the magnetic field is able to stabilise the MRI only above $\zeta_B$, the non-turbulent surface layer is inefficient at reducing the diffusion.

All these results can be explained using the vertical averaging procedure described in Section~\ref{sec:induction_average}. The relevant average resistivity determining the diffusion of the magnetic field is the inverse of the volume average conductivity performed up to the height $\zeta_B$ (where the magnetic pressure equals the thermal pressure). The averaged radial velocity relevant to the advection of magnetic flux is conductivity weighted (also performed up to the height $\zeta_B$) and can therefore be very different from the density-weighted average that determines the transport of mass \citep{ogilvie01}.

\subsection{Status of the `magnetic flux problem' and future directions}
Early theoretical calculations predicted that the magnetic field diffuses faster than it is advected by the accretion flow, implying that a significant magnetic flux should not be present in thin accretion discs \citep{lubow94a,heyvaerts96}. On the other hand, observations show the existence of strong outflows in a variety of objects containing an accretion disc (AGN, XRB, cataclysmic variables, T-Tauri stars). Theoretical models explaining these outflows generally need a strong net vertical magnetic field to accelerate these outflows through the magnetorotational mechanism \citep{blandford82} or the Blandford--Znajek mechanism for systems containing a black hole \citep{blandford77}.  How could such strong vertical magnetic fields be obtained?

It has been proposed that the presence of a non-turbulent layer at the surface of the disc could solve the problem by drastically reducing the diffusion of the magnetic field \citep{bisnovatyi-kogan07,bisnovatyi-kogan12,rothstein08,lovelace09}. We have studied this possibility in Section~\ref{sec:lovelace} and found that the non-turbulent surface layer is inefficient at reducing the diffusion if it lies where the large scale magnetic field is dominant (as it is expected to).  This conclusion should ideally be verified using numerical simulations of turbulent discs.

In Paper~I \citep{guilet12}, we showed that the situation was not that bad for weak magnetic fields ($\beta_0\gg 1$). Indeed, the vertical structure of the velocity and magnetic field induce a faster advection and a slower diffusion of the magnetic flux as compared to the vertical averaging used before by \citet{lubow94a} and others. As a consequence the inclination of the magnetic field lines in a stationary situation, which is regulated by the relative efficiency of advection and diffusion processes, could be 10 to 50 times larger than thought before, for weak enough magnetic fields. This encouraging result was obtained in the idealised case of uniform diffusion coefficients. In this paper, we have confirmed the robustness of this result with different vertical profiles of the diffusion coefficients, modelling either a fully turbulent disc or one containing a non-turbulent dead zone near its midplane.  

Though encouraging, these new results are so far insufficient for very thin discs, and obtaining strong magnetic fields here remains problematic. Additional effects neglected in the present study could help further the advection of the magnetic flux: 
\begin{itemize}
\item A more extended corona than the present isothermal model (either because it is hot or because of magnetic support) would further reduce the diffusion (by increasing $\zeta_B$ for a given magnetic field strength) and possibly also increase the advection efficiency. 
\item An anisotropic diffusion could also contribute to reducing the diffusion of magnetic flux: if the vertical diffusion of the magnetic field is slower than the radial diffusion of angular momentum by some numerical factor, this would favour the advection of magnetic flux over diffusion. Such an anisotropy of the diffusion processes is suggested by the numerical results of \citet{lesur09}, though more numerical studies are certainly necessary to confirm this (the vertical diffusion of a radial magnetic field has never been measured so far, since \citet{lesur09} only measured the vertical diffusion of an azimuthal magnetic field). 
\item Also we note that the diffusion properties of the magnetically dominated corona have never been investigated. The importance of this region of the disc for the transport of magnetic flux has been highlighted by this paper and Paper~I as well as by \citet{beckwith09}. If this region of the disc was not efficient at diffusing magnetic field (though it is highly dynamic), this could help further reduce the diffusion. 
\end{itemize}

Even with these additional effects something more may be needed to achieve strong magnetic field that can launch a strong outflow. In Paper~I, we proposed that this additional process could be the outflow itself if some outflow can be obtained for rather weak magnetic field fields. In this scenario the advection due to the effective turbulent viscosity would be efficient up to a certain strength of order $\beta\sim 10^2-10^4$ (depending on the aspect ratio of the disc). A (weak) outflow would then be launched and would enable further advection of the magnetic flux, finally reaching strong magnetic fields. The effect of an outflow on the advection of magnetic flux has been considered by \cite{lovelace09,bisnovatyi-kogan12,guilet12}. A few numerical simulations suggest the possibility of such outflows with a weak magnetic field. \citet{murphy10} have obtained a magneto-centrifugal outflow from a weakly magnetised disc with $\beta \sim 100-1000$. An outflow was also obtained in the local shearing box model for even weaker magnetic fields ($\beta=10^4-10^8$) by \citet{suzuki09} and \citet{okuzumi11}.  Much more work is, however, needed to understand the properties of outflows that may be launched from weakly magnetised discs and the effect of such outflows on the transport of magnetic flux.

We found in Paper~I that the aforementioned scenario to obtain strong magnetic fields by inward flux transport could operate in discs with an aspect ratio above a (rather uncertain) critical value. This criterion may be met in protoplanetary discs, the aspect ratio of which is not very small: $H/r \sim 0.05-0.1$. Strong large-scale poloidal magnetic fields may then be obtained, enabling the launching of a magnetocentrifugal outflow which could explain the jets observed in T-Tauri stars as argued by \citet{ferreira06}. In the soft state of X-ray binaries, the accretion disc is thinner (unless the luminosity approaches the Eddington luminosity), and is therefore likely to be too thin for the above scenario to operate. The disc would then be threaded by a weak magnetic field, which is not able to drive a powerful jet. This may explain the absence of jets in the soft state of X-ray binaries \citep{fender99,coriat11}. In the hard state on the other hand, the inner part of the accretion flow is thought to be radiatively inefficient and geometrically thick. In principle, a strong magnetic field could therefore accumulate in the inner geometrically thick disc, and drive a powerful jet such as simulated e.g. by \citet{tchekhovskoy11}.  In order to reach this inner region, however, the magnetic flux also needs to be transported inwards in the outer parts of the disc, which may be geometrically thin. The results of this paper and Paper~I show that this flux transport is possible for weak enough magnetic fields even if the disc is thin. One could therefore envision a situation where the outer thin accretion is weakly magnetised but allows magnetic flux to be transported inwards. This magnetic flux can then accumulate and become dynamically significant in the inner geometrically thick accretion flow, where a strong jet is launched. This idea should be tested in a global disc model, in which the time-evolution of the magnetic flux could be computed using the transport rates computed in this paper.

\section*{Acknowledgments}
This research was supported by STFC.  We thank Henk Spruit, Henrik Latter, Cl\'ement Baruteau, and S\'ebastien Fromang for useful discussions.

\bibliography{discs}

\begin{thebibliography}{}

\bibitem[\protect\citeauthoryear{{Balbus} \& {Hawley}}{{Balbus} \&
  {Hawley}}{1991}]{balbus91}
{Balbus} S.~A.,  {Hawley} J.~F.,  1991, \apj, 376, 214

\bibitem[\protect\citeauthoryear{{Beckwith}, {Hawley} \& {Krolik}}{{Beckwith}
  et~al.}{2009}]{beckwith09}
{Beckwith} K.,  {Hawley} J.~F.,    {Krolik} J.~H.,  2009, \apj, 707, 428

\bibitem[\protect\citeauthoryear{{Bisnovatyi-Kogan} \&
  {Lovelace}}{{Bisnovatyi-Kogan} \& {Lovelace}}{2007}]{bisnovatyi-kogan07}
{Bisnovatyi-Kogan} G.~S.,  {Lovelace} R.~V.~E.,  2007, \apjl, 667, L167

\bibitem[\protect\citeauthoryear{{Bisnovatyi-Kogan} \&
  {Lovelace}}{{Bisnovatyi-Kogan} \& {Lovelace}}{2012}]{bisnovatyi-kogan12}
{Bisnovatyi-Kogan} G.~S.,  {Lovelace} R.~V.~E.,  2012, \apj, 750, 109

\bibitem[\protect\citeauthoryear{{Blandford} \& {Payne}}{{Blandford} \&
  {Payne}}{1982}]{blandford82}
{Blandford} R.~D.,  {Payne} D.~G.,  1982, \mnras, 199, 883

\bibitem[\protect\citeauthoryear{{Blandford} \& {Znajek}}{{Blandford} \&
  {Znajek}}{1977}]{blandford77}
{Blandford} R.~D.,  {Znajek} R.~L.,  1977, \mnras, 179, 433

\bibitem[\protect\citeauthoryear{{Coriat}, {Corbel}, {Prat}, {Miller-Jones},
  {Cseh}, {Tzioumis}, {Brocksopp}, {Rodriguez}, {Fender} \&
  {Sivakoff}}{{Coriat} et~al.}{2011}]{coriat11}
{Coriat} M.,  {Corbel} S.,  {Prat} L.,  {Miller-Jones} J.~C.~A.,  {Cseh} D.,
  {Tzioumis} A.~K.,  {Brocksopp} C.,  {Rodriguez} J.,  {Fender} R.~P.,
  {Sivakoff} G.~R.,  2011, \mnras, 414, 677

\bibitem[\protect\citeauthoryear{{Fender}, {Corbel}, {Tzioumis}, {McIntyre},
  {Campbell-Wilson}, {Nowak}, {Sood}, {Hunstead}, {Harmon}, {Durouchoux} \&
  {Heindl}}{{Fender} et~al.}{1999}]{fender99}
{Fender} R.,  {Corbel} S.,  {Tzioumis} T.,  {McIntyre} V.,  {Campbell-Wilson}
  D.,  {Nowak} M.,  {Sood} R.,  {Hunstead} R.,  {Harmon} A.,  {Durouchoux} P.,
    {Heindl} W.,  1999, \apjl, 519, L165

\bibitem[\protect\citeauthoryear{{Ferreira}, {Dougados} \& {Cabrit}}{{Ferreira}
  et~al.}{2006}]{ferreira06}
{Ferreira} J.,  {Dougados} C.,    {Cabrit} S.,  2006, \aap, 453, 785

\bibitem[\protect\citeauthoryear{{Fleming} \& {Stone}}{{Fleming} \&
  {Stone}}{2003}]{fleming03}
{Fleming} T.,  {Stone} J.~M.,  2003, \apj, 585, 908

\bibitem[\protect\citeauthoryear{{Fromang}, {Hennebelle} \&
  {Teyssier}}{{Fromang} et~al.}{2006}]{fromang06}
{Fromang} S.,  {Hennebelle} P.,    {Teyssier} R.,  2006, \aap, 457, 371

\bibitem[\protect\citeauthoryear{{Fromang} \& {Nelson}}{{Fromang} \&
  {Nelson}}{2006}]{fromang06b}
{Fromang} S.,  {Nelson} R.~P.,  2006, \aap, 457, 343

\bibitem[\protect\citeauthoryear{{Fromang}, {Terquem} \& {Balbus}}{{Fromang}
  et~al.}{2002}]{fromang02}
{Fromang} S.,  {Terquem} C.,    {Balbus} S.~A.,  2002, \mnras, 329, 18

\bibitem[\protect\citeauthoryear{{Fromang}, {Terquem} \& {Nelson}}{{Fromang}
  et~al.}{2005}]{fromang05}
{Fromang} S.,  {Terquem} C.,    {Nelson} R.~P.,  2005, \mnras, 363, 943

\bibitem[\protect\citeauthoryear{{Gammie}}{{Gammie}}{1996}]{gammie96}
{Gammie} C.~F.,  1996, \apj, 457, 355

\bibitem[\protect\citeauthoryear{{Guilet} \& {Ogilvie}}{{Guilet} \&
  {Ogilvie}}{2012}]{guilet12}
{Guilet} J.,  {Ogilvie} G.~I.,  2012, \mnras, 424, 2097

\bibitem[\protect\citeauthoryear{{Heyvaerts}, {Priest} \& {Bardou}}{{Heyvaerts}
  et~al.}{1996}]{heyvaerts96}
{Heyvaerts} J.,  {Priest} E.~R.,    {Bardou} A.,  1996, \apj, 473, 403

\bibitem[\protect\citeauthoryear{{Kley} \& {Lin}}{{Kley} \&
  {Lin}}{1992}]{kley92}
{Kley} W.,  {Lin} D.~N.~C.,  1992, \apj, 397, 600

\bibitem[\protect\citeauthoryear{{Lesur} \& {Longaretti}}{{Lesur} \&
  {Longaretti}}{2009}]{lesur09}
{Lesur} G.,  {Longaretti} P.-Y.,  2009, \aap, 504, 309

\bibitem[\protect\citeauthoryear{{Lovelace}, {Rothstein} \&
  {Bisnovatyi-Kogan}}{{Lovelace} et~al.}{2009}]{lovelace09}
{Lovelace} R.~V.~E.,  {Rothstein} D.~M.,    {Bisnovatyi-Kogan} G.~S.,  2009,
  \apj, 701, 885

\bibitem[\protect\citeauthoryear{{Lubow}, {Papaloizou} \& {Pringle}}{{Lubow}
  et~al.}{1994}]{lubow94a}
{Lubow} S.~H.,  {Papaloizou} J.~C.~B.,    {Pringle} J.~E.,  1994, \mnras, 267,
  235

\bibitem[\protect\citeauthoryear{{Miller} \& {Stone}}{{Miller} \&
  {Stone}}{2000}]{miller00}
{Miller} K.~A.,  {Stone} J.~M.,  2000, \apj, 534, 398

\bibitem[\protect\citeauthoryear{{Murphy}, {Ferreira} \& {Zanni}}{{Murphy}
  et~al.}{2010}]{murphy10}
{Murphy} G.~C.,  {Ferreira} J.,    {Zanni} C.,  2010, \aap, 512, A82

\bibitem[\protect\citeauthoryear{{Ogilvie} \& {Livio}}{{Ogilvie} \&
  {Livio}}{2001}]{ogilvie01}
{Ogilvie} G.~I.,  {Livio} M.,  2001, \apj, 553, 158

\bibitem[\protect\citeauthoryear{{Okuzumi} \& {Hirose}}{{Okuzumi} \&
  {Hirose}}{2011}]{okuzumi11}
{Okuzumi} S.,  {Hirose} S.,  2011, \apj, 742, 65

\bibitem[\protect\citeauthoryear{{Rothstein} \& {Lovelace}}{{Rothstein} \&
  {Lovelace}}{2008}]{rothstein08}
{Rothstein} D.~M.,  {Lovelace} R.~V.~E.,  2008, \apj, 677, 1221

\bibitem[\protect\citeauthoryear{{Spruit}}{{Spruit}}{2010}]{spruit10}
{Spruit} H.~C.,  2010, in {T.~Belloni} ed., Lecture Notes in Physics, Berlin
  Springer Verlag Vol.~794 of Lecture Notes in Physics, Berlin Springer Verlag,
  {Theory of Magnetically Powered Jets}.
p.~233

\bibitem[\protect\citeauthoryear{{Suzuki} \& {Inutsuka}}{{Suzuki} \&
  {Inutsuka}}{2009}]{suzuki09}
{Suzuki} T.~K.,  {Inutsuka} S.-i.,  2009, \apjl, 691, L49

\bibitem[\protect\citeauthoryear{{Takeuchi} \& {Lin}}{{Takeuchi} \&
  {Lin}}{2002}]{takeuchi02}
{Takeuchi} T.,  {Lin} D.~N.~C.,  2002, \apj, 581, 1344

\bibitem[\protect\citeauthoryear{{Tchekhovskoy}, {Narayan} \&
  {McKinney}}{{Tchekhovskoy} et~al.}{2011}]{tchekhovskoy11}
{Tchekhovskoy} A.,  {Narayan} R.,    {McKinney} J.~C.,  2011, \mnras, 418, L79

\bibitem[\protect\citeauthoryear{{Terquem}}{{Terquem}}{2003}]{terquem03}
{Terquem} C.~E.~J.~M.~L.~J.,  2003, \mnras, 341, 1157

\end{thebibliography}

\bsp
\label{lastpage}

\end{document}